 \documentclass[acmsmall]{acmart}
\AtBeginDocument{%
  \providecommand\BibTeX{{%
    \normalfont B\kern-0.5em{\scshape i\kern-0.25em b}\kern-0.8em\TeX}}}

\setcopyright{acmcopyright}
\copyrightyear{2023}
\acmYear{2023}
\acmDOI{XXXXXXX.XXXXXXX}

\acmConference[CSCW '24]{ACM SIGCHI Conference on Computer-Supported Cooperative Work & Social Computing (CSCW) }{November 9-13, 2024}{San José, Costa Rica}
\acmPrice{15.00}
\acmISBN{978-1-4503-XXXX-X/18/06}
\usepackage{balance}       % to better equalize the last page
\usepackage{graphics}      % for EPS, load graphicx instead 
\usepackage{hyperref}
\usepackage{color}
\usepackage{booktabs}
\usepackage{textcomp}
\usepackage{subcaption}
\usepackage{enumerate}
\usepackage{xcolor}
\usepackage{lipsum}% http://ctan.org/pkg/lipsum
\usepackage{makecell}
\usepackage{multicol}
\usepackage{multirow}
\usepackage{array}
\usepackage{verbatimbox}
\usepackage{enumitem}
\usepackage{amsmath}
\usepackage{stfloats}
\usepackage{graphicx}
\usepackage{amsthm}
\usepackage{listings}
\usepackage{caption} 
\usepackage[export]{adjustbox}
\usepackage{xspace}
\usepackage{epsfig}
\usepackage[linesnumbered]{algorithm2e}
\usepackage{algpseudocode}
\usepackage{tabularx}
\usepackage{arydshln}
\usepackage[bottom]{footmisc}
\usepackage{tcolorbox}
\usepackage{stackengine}
\usepackage{placeins}

\newcommand*{\ie}{\textit{i.e.},\xspace}

\newcolumntype{L}[1]{>{\raggedright\let\newline\\\arraybackslash\hspace{0pt}}m{#1}}
\newcolumntype{C}[1]{>{\centering\let\newline\\\arraybackslash\hspace{0pt}}m{#1}}
\newcolumntype{R}[1]{>{\raggedleft\let\newline\\\arraybackslash\hspace{0pt}}m{#1}}

\definecolor{ProBlue}{HTML}{DAF3F7}
\definecolor{CPPink}{HTML}{F9E3E3}
\newcommand\CP[1]{\colorbox{CPPink}{CP#1}}
\newcommand\Pro[1]{\colorbox{ProBlue}{Pro#1}}
\newcommand\pquote[2]{{``\textit{#2}'' (\colorbox{ProBlue}{Pro#1})}}
\newcommand\epquote[2]{{``\textit{#2}'' (\colorbox{CPPink}{CP#1})}}

\makeatletter
\def\thickhline{%
  \noalign{\ifnum0=`}\fi\hrule \@height \thickarrayrulewidth \futurelet
   \reserved@a\@xthickhline}
\def\@xthickhline{\ifx\reserved@a\thickhline
               \vskip\doublerulesep
               \vskip-\thickarrayrulewidth
             \fi
      \ifnum0=`{\fi}}
\makeatother

\makeatletter
\def\thickhlinespace{%
  \addlinespace[1ex]
  \noalign{\ifnum0=`}\fi\hrule \@height \thickarrayrulewidth \futurelet
   \reserved@a\@xthickhline
   \addlinespace[1ex]
   }
\def\@xthickhlinespace{\ifx\reserved@a\thickhline
               \vskip\doublerulesep
               \vskip-\thickarrayrulewidth
             \fi
      \ifnum0=`{\fi}}
\makeatother

\newlength{\thickarrayrulewidth}
\newlength\Origarrayrulewidth

\algnewcommand{\IfThenElse}[3]{% \IfThenElse{<if>}{<then>}{<else>}
  \State \algorithmicif\ #1\ \algorithmicthen\ #2\ \algorithmicelse\ #3}

\definecolor{downredcolor}{HTML}{e31a1c}
\definecolor{upgreencolor}{HTML}{33a02c}

\definecolor{DarkGreen}{HTML}{5DAC81}
\acmSubmissionID{7482}

\begin{document}

%%
%% The "title" command has an optional parameter,
%% allowing the author to define a "short title" to be used in page headers.
% \title[Post-Treatment Cancer Care]{``They could be dying, but they don't call'': Understanding Challenges and AI Potential in Patient-Provider Communication After Cancer Treatments}

% If it asked questions it would have noticed the symptoms.
% \title[``An AI Would Have Noticed That Symptom'']{``An AI Would Have Noticed That Symptom'': Understanding Challenges and AI Potential in Patient-Provider Communication After Cancer Treatments}

\title[``I Wish There Were an AI'']{``I Wish There Were an AI'': Challenges and AI Potential in Cancer Patient-Provider Communication
}

%%
%% The "author" command and its associated commands are used to define
%% the authors and their affiliations.
%% Of note is the shared affiliation of the first two authors, and the
%% "authornote" and "authornotemark" commands
%% used to denote shared contribution to the research.
\author{Ziqi Yang}
% \authornote{Both authors contributed equally to this research.}
\email{ziq.yang@northeastern.edu}
\orcid{1234-5678-9012}
\affiliation{%
  \institution{Northeastern University}
  \streetaddress{440 Huntington Ave}
  \city{Boston}
  \state{Massachusetts}
  \country{USA}
  \postcode{01228}
}

\author{Xuhai Xu}
\affiliation{%
  \institution{Massachusetts Institutte of Technology}
  \city{Boston}
  \state{Massachusetts}
  \country{USA}
  }

\author{Bingsheng Yao}
\affiliation{
  \institution{Northeastern University}
  \city{Boston}
  \state{Massachusetts}
  \country{USA}
}

\author{Jiachen Li}
\affiliation{
  \institution{Northeastern University}
  \city{Boston}
  \state{Massachusetts}
  \country{USA}
}

\author{Jennifer Bagdasarian}
\email{Jbagdas1@jh.edu}
\affiliation{%
  \institution{Johns Hopkins University}
  % \city{Baltimore}
  % \state{Maryland}
  \country{USA}
}
\author{Guodong (Gordon) Gao}
\email{gordon.gao@jhu.edu}
\affiliation{%
  \institution{Johns Hopkins University}
  % \city{Baltimore}
  % \state{Maryland}
  \country{USA}
}

\author{Dakuo Wang}
\authornote{Corresponding author d.wang@northeastern.edu}
\email{d.wang@northeastern.edu}
\affiliation{%
  \institution{Northeastern University}
  % \city{Boston}
  % \state{Massachusetts}
  \country{USA}
}
%%
%% By default, the full list of authors will be used in the page
%% headers. Often, this list is too long, and will overlap
%% other information printed in the page headers. This command allows
%% the author to define a more concise list
%% of authors' names for this purpose.
\renewcommand{\shortauthors}{Yang, et al.}

%%
%% The abstract is a short summary of the work to be presented in the
%% article.
\begin{abstract}
Patient-provider communication has been crucial to cancer patients' survival after their cancer treatments. 
However, the research community and patients themselves often overlook the communication challenges after cancer treatments as they are overshadowed by the severity of the patient's illness and the variety and rarity of the cancer disease itself. 
Meanwhile, the recent technical advances in AI, especially in Large Language Models (LLMs) with versatile natural language interpretation and generation ability, demonstrate great potential to support communication in complex real-world medical situations.
By interviewing six healthcare providers and eight cancer patients, our goal is to explore the providers' and patients' communication barriers in the post-cancer treatment recovery period, their expectations for future communication technologies, and the potential of AI technologies in this context.
Our findings reveal several challenges in current patient-provider communication, including the knowledge and timing gaps between cancer patients and providers, their collaboration obstacles, and resource limitations.
Moreover, based on providers' and patients' needs and expectations, we summarize a set of design implications for intelligent communication systems, especially with the power of LLMs.
Our work sheds light on the design of future AI-powered systems for patient-provider communication under high-stake and high-uncertainty situations.
% Thus, we provide design implications for future AI-powered systems to support their communication.
\end{abstract}

% Link to figures: https://www.figma.com/file/6AJL3jabGELgtUZaflQp5y/CSCW-Cancer-Care-Paper?type=design&node-id=0%3A1&mode=design&t=jmTUyvRpY5VEVZ3c-1

%%
%% The code below is generated by the tool at http://dl.acm.org/ccs.cfm.
%% Please copy and paste the code instead of the example below.
%%



\maketitle

\section{Introduction}
% Overall structure
% 1. cancer has unique difficulty, commmunication solve
% 2. existing communication can not solve difficulty
% 3. recent advance of AI shed light on solving, but how to better utilize ...to solve ... has not been 
% 4. thus, RQ: 1. practice, need, challenge 2. how AI solve

% threatening after surgery, have a period of 

% still dangerous, because (high level)
% 1. vulnerable (severe)
% 2. rare
% 3. emotional

% -> difficulty, is my motivation
% uniqueness of cancer

% lack of communication -> can not rescue (use some existing work, solution highly depends on in-time communication) or better communication can improve survival (good)

% although some research supports communication, hypothesis 1: but is not effective for cancer patients
% e.g. cancer limited knowledge, tech can not educate; not able to use technology

% a post-treatment system specifically for cancer...

% 1. find specific challenges
% 2. how can AI can address cancer issues
% result: solve the three difficulty
% approach: facilitate communication

Once cancer patients are discharged from hospitals after treatment, they are still under significant health risks, such as post-treatment complications and cancer recurrence~\cite{wang2018prevalence, jones2016cancer, offiah2011post, mahvi2018local, ferlay2021cancer}. 
With the rarity of cancer, patients generally have limited knowledge and a high level of stress about their disease~\cite{siegel2021cancer, gupta2015review, akram2017awareness, mosher2016mental}. 
Moreover, post-treatment conditions and recovery strategies for different types of cancers vary significantly across individuals % so patients may be confused whether a symptom is considered abnormal
~\cite{ilic2016epidemiology, heng2011evolutionary, iqbal2015differences}.
Therefore, they may not be aware of all abnormalities of their body and inform their healthcare providers in time ~\cite{gupta2015review, akram2017awareness}. 

Prior studies show that successful patient-provider communication could improve the survival of cancer patients, as the communication can lead to tailored provider instructions and more timely interventions~\cite{frenkel2016exceptional}.
Synchronous communication channels, such as phone calls, are effective and preferable, but they are not sustainable due to the limited bandwidth of healthcare providers and the uncertain availability of patients.
Thus, researchers have proposed various asynchronous communication technologies to support patient-provider communication in post-treatment cancer care practices~\cite{yu2021deep, penedo2020increasing}.
% Ensuring patients' health after cancer treatments has been challenging for cancer patients and their providers. 
% For example, patients may experience post-treatment complications, which require immediate support from healthcare providers \cite{}. Therefore, it is important for cancer care providers to follow-up on patients' health conditions and be informed of any abnormalities~\cite{}. In such cases, cancer providers could take action to readmit patients to hospitals or offer instructions remotely to tend to patients' conditions. Thus, effective post-treatment patient-provider communication could  promote cancer patients' survival~\cite{}. Currently, telehealth technologies have been widely utilized by healthcare providers to communicate with patients at home. 
% However, these existing practices do not prove to be effective for cancer patients' urgent needs and severe symptoms, resulting in high readmission and mortality rates~\cite{}.
% The critical question of how technologies can better support patient-provider communication for cancer patients after treatments remains unanswered.
% For example,
Existing research has explored a variety of technologies, such as instant messaging, mobile health application (mHealth) tools, and online forums, to support asynchronous patient-provider communication~\cite{andy2021understanding, khurana2019doctor, goncalves-bradley_mobile_2020, donelan_patient_2019}. These tools have facilitated communication by enabling remote communication, enriching personal health information with pictures and tracking devices, and offering peer support ~\cite{goncalves-bradley_mobile_2020, andy2021understanding}.
However, these technologies are mainly designed for general communication needs in non-urgent scenarios or widely known chronic diseases. They may not work for post-cancer treatment, as patient symptoms can be highly personalized. 
Often the case, cancer patients have questions about the existing medical advice given their specific situations. Since they can only find limited information through mHealth tools or forums, they would have to contact their providers and wait for responses, which could cause undesirable delays, especially when asynchronous communication requires clarification or education with multiple back-and-forth rounds.
% and thus may not be able to address the challenges for cancer patients after their cancer treatments.
% This gap calls for novel systems to support patient-provider communication, specifically for cancer patients and their providers.

The recent technological boost of AI reveals the promising possibility of supporting cancer patients' treatments and healthcare. In particular, Large-Language Models (LLMs) have shown the capability to collect personal health information with Conversational Agents (CAs), provide mental support, integrate domain knowledge, and assist clinical workers~\cite{ma2023understanding,lee2020biobert,yang_integrating_2023,ali2023using}. However, the question of how AI could support patient-provider communication in post-treatment cancer care, especially due to its rarity, severity, and variety, is under-explored.

In this paper, we aim to study the communication challenges in post-treatment cancer care and explore the potential for AI to address these challenges.
We interviewed six healthcare providers specializing in cancer and eight cancer patients who have gone through cancer treatments to understand their communication experiences and expectations for future technologies.
We inquired about their current practices and challenges in communicating cancer-related information after cancer treatments.
We then encouraged cancer patients and providers to depict how they would like AI technologies to support them during communication.

We discovered that the patient-provider communication challenges in post-treatment cancer care mainly result from the knowledge gap and timing gap between cancer patients and their healthcare providers.
Our findings indicate the potential of leveraging AI to fill in these gaps. For patients, AI can explain provider instructions and help patients reflect on symptoms. For providers, AI can help annotate patient health information.
In addition, besides the patient's directly responsible provider (\ie cancer-specific), AI can assist the patient in navigating the patient's cancer support network and collaborating with other providers for post-treatment healthcare. 
On the other hand, despite stakeholders' expectations for tailoring the AI system to each patient, the concern about system reliability and privacy remains.
We summarize our findings into a set of design implications for AI to empower cancer patient-provider communication through conversational agents.% and connecting the cancer support network.

Our work makes contributions in the following aspects:
\begin{itemize}
    \item We identify key patient-provider asynchronous communication challenges in post-treatment cancer care for cancer patients and discuss the causes that lead to those challenges.
    \item Our findings reveal the potential for AI to support patient-provider communication in the context of high-risk diseases.
    \item We provide design implications for AI-powered systems to support patient-provider communication in cancer patients' post-treatment healthcare.
\end{itemize}

% Thus, we inform system designers and researchers of challenges in patient-provider communication in post-treatment cancer care, and what strategies and guidelines might be for AI technologies to provide such communication support. We also reveal the potential for AI to support patient-provider communication in a highly risky disease and post-surgery context. Eventually, our findings could benefit patients, providers, and family members by instructing the design and development of novel communication technologies.

\section{Related Work}
\subsection{Post-Treatment Cancer Care}
% conclusion: the challenge of cancer in the intro; better communication can 
% cancer severity-related topic

Prior studies have explored challenges for different stakeholders in cancer care practices~\cite{parkerBreastCancerUnique2009,thorne2013communication,surboneNewChallengesCommunication2012,penedo2020increasing}. In works focusing on post-surgery cancer care, researchers identified the challenges for patients in adapting to surgery outcomes and complications\cite{silva_ostomy_2020, mosher2016mental}, as well as the need for family support and patient education ~\cite{ran_quality_2016}. Healthcare providers experience significant stress and thus require social support and reduced workload ~\cite{vachon1995staff, surboneNewChallengesCommunication2012}. The work for caregivers at home is physically, emotionally, socially, and financially demanding ~\cite{kent2016caring}. Specifically, in patient-provider communication practices, the stakeholders experience gaps in knowledge and medical resources.  Many cancer patients experience mental issues such as depression, and thus, researchers have proposed design frameworks for technologies to support their mental health~\cite{suh_parallel_2020}.  Recent research also studied communication strategies for parents (caregivers) in communicating with their child (patient) in pediatric cancer care, focusing on children's emotions ~\cite{seo2021challenges}. 

However, most of the qualitative studies focus on the in-person communication and mental aspects between cancer patients and providers, which takes place before or during patients' cancer treatments~\cite{kabir2019m,jacobs2014cancer}. Although some studies explored cancer patients' challenges and offered information support for cancer patients with mHealth applications, the engagement of healthcare providers in cancer care communication is limited~\cite{jacobsMyPathInvestigatingBreast2018}. After cancer patients are discharged from the hospital, it is unclear what are the major communication practices, needs and challenges patients and providers encounter.

\subsection{Patient-Provider Communication Technologies}

The use of digital information and communication technologies for remote health service is defined as telehealth~\cite{noauthor_telehealth_nodate}. This includes mobile health technologies(mHealth) ~\cite{goncalves-bradley_mobile_2020}, virtual visits (teleconsultation)~\cite{chandwani_stitching_2018, bhat_infrastructuring_2021}and TeleMonitoring~\cite{serban_i_2023} These tools have lifted practitioner and patient burdens in communication. Some medical work quantitatively reviewed telehealth use in cancer and post-surgery care ~\cite{hong_digital_2020, donelan_patient_2019}. 
Despite their effectiveness in connecting patients and providers~\cite{clark_understanding_2021,bhat_infrastructuring_2021}, many still refer to traditional methods such as phone calls, emails, or text messaging for asynchronous communication due to the poor system integration, cost concerns, and usability~\cite{chandwani_stitching_2018, ye_e-mail_2010}.
Given these challenges, the HCI community has explored technologies to support asynchronous patient-provider communication. Earlier this decade, HCI researchers designed mHealth applications and tracking devices to collect patient health information for collaborative review ~\cite{chung_more_2015, schroeder_supporting_2017, jacobs_comparing_2015, raj_understanding_2017}, yet there was a lack of multitude on the data provided~\cite{pine_data_2018}. Later, some studies revealed the advantages and challenges of using messaging apps for remote patient-provider communication in China and India~\cite{wang_please_2020, karusala_making_2020}. 
% For example, \citet{wang_please_2020}'s work in 2020 studied the use of WeChat, an instant messaging application, to support patient care in clinics in China, pointing out the multitude of information and the burden for providers. However, these apps were used as complementary solutions, which exert privacy and security concerns ~\cite{wang_please_2020}.
% However, no recent work explored or evaluated how the design of these applications, provides information, manages or fails to support patient care. Therefore, we aim to study current practices of remote patient-provider communication in post-surgery cancer care contexts.
However, most of the communication technologies are designed for the general public, while cancer patients' communication needs introduce more complexity, urgency, and severity~\cite{chavarri2021providing}. Thus, these technologies may not be applicable for patient-provider communication in post-treatment cancer care. In this study, we aim to explore the challenges for cancer patients using the existing technologies, and how we should design future technologies to support their patient-provider communication.

\subsection{AI for Healthcare Communication}

In recent years, the development of AI has shed light on supporting clinical work and healthcare communication. Language models have been specifically trained with medical knowledge and data to serve the healthcare community better, such as Med-PaLM2 (based on Google PaLM2)\cite{anil2023palm}, BioGPT~\cite{luo2022biogpt}, BioBERT~\cite{lee2020biobert}, UmlsBERT~\cite{michalopoulos_umlsbert_2020}, SciBERT~\cite{beltagy2019scibert}.
Meanwhile, a significant amount of work has examined and tested the use of ChatGPT\footnote{https://chat.openai.com} in medical and healthcare scenarios, including answering medical questions~\cite{johnson2023assessing, gilson2023does}, generating summaries~\cite{ali2023using,liu2023utility}, clinical text mining ~\cite{tang_does_2023}, or facilitating the entire clinical workflow ~\cite{rao_assessing_2023,liu2023utility}. For example, some recent work designed and evaluated a method to augment ChatGPT with UMLS~\cite{yang_integrating_2023}. For cancer care practices, researchers also look for factors that contribute to survival, such as predicting hospital readmissions and the long-term survival of cancer patients~\cite{masum2021data, masum2022data, mohanty2022machine}, or used deep-learning methods to support diagnosis with cancer imaging  ~\cite{yu2021deep}. These technical advances motivate us to leverage AI strengths in medical knowledge to support healthcare work in complicated scenarios like cancer care.

Some recent work in the human-computer interaction domain has utilized Language Models for healthcare practices, revealing opportunities for LLMs to promote patient-provider communication. Recent research has explored the use of LLM-powered chatbots for mental health~\cite{wu_mindshift_2023}, patient communication\cite{fang_socializechat_2023,seo_chacha_2023}, or patient journaling\cite{kim_mindfuldiary_2023}. 
% In 2023, \citet{seo_chacha_2023} designed and developed an LLM-powered mobile chatbot for children to better express their emotions. \citet{fang2023socializechat} used GPT-4 to build an Augmentative and Alternative Communication (AAC) tool, SocializeChat, to help patients with motor or speech impairments communicate with others through gaze inputs. 
%A brief user test showed that it was effective in delivering content preferences and communication style, which may potentially promote patients' social closeness. 
For mental support, recent research has used LLMs to facilitate psychiatric patients' journaling, encourage emotional expression, or provide mental health interventions~\cite{kim_mindfuldiary_2023, seo_chacha_2023, wu_mindshift_2023}.  In addition, some related work has explored conversational agents to support communication between multiple stakeholders in healthcare work, such as using LLM to simulate patients and providers and multi-agent debate~\cite{chen_llm-empowered_2023,chan_chateval_2023}.
% For example, \citet{shin2023introbot}'s study used NLP methods to implement a chatbot, IntroBot to facilitate online group discussion by generating keywords. The results show that IntroBot helped to establish higher levels of trust, cohesion, and interaction quality, as well as generated more ideas in a collaborative brainstorming task. 
% In a recent work by \citet{}, researchers used different versions of LLM to simulate both the patient and the healthcare provider in outpatient sessions by designing and iterating prompts. \citet{} studied the feasibility and dynamics of using LLM for . 

Despite the AI advances in supporting healthcare communication, there is a lack of work on how AI could support patient-provider communication in post-treatment cancer care, especially how AI could address the challenges due to rarity, peril, and variety of cancer. Thus, we aim to explore the AI potential through our study with post-treatment cancer patients and providers.
% However, most of the current work focuses on the user in patient personal health, and little work has explored the use of LLM-powered conversational agents (CAs) for healthcare providers or patient-provider communication. ~\cite{} (the previous article) Yet it is still unclear whether these CAs could be of use in the context of clinical practices. Therefore, we aim to explore the opportunities for LLMs in clinical practices, such as cancer or post-treatment care.
% - novel interactions
% - CA for medical/healthcare
% - ...

% could add more lit review

% future: more advanced methods in NLP, possibility
\section{Methodology}

To understand current asynchronous communication practices, needs, and challenges between cancer patients and  their healthcare providers, we conducted semi-structured interviews with six healthcare providers and eight cancer patients following the research questions: 
\begin{enumerate}
    % \item What are the current communication practices and needs for patients and their providers after the cancer treatments?
    \item RQ1: What are the communication challenges in the patient-provider communication for cancer patients after the cancer treatments?
    \item RQ2: How should we design technologies leveraging technical advances in AI to support patient-provider communication in post-treatment cancer care?
\end{enumerate}

\renewcommand{\arraystretch}{1.3}
\label{methods:provider-demographics}
\begin{table}[t]
    \centering
    \resizebox{1\linewidth}{!}{
    \begin{tabular}{c|c|c|c|c}
    % \toprule
    \hline \hline
        \textbf{Pro\#} & \textbf{Gender} &\textbf{Role} & \textbf{Expertise} & \textbf{Years of experience} \\
        \hline
        Pro1 & Female & Nurse Practitioner& Major surgery, orthopedic cancer & 10-20 years\\ 
        Pro2 & Male & Surgeon& Minimally invasive surgery, GI cancer & 5-10 years \\ 
        Pro3 & Female &  Surgeon & GI cancer & 5-10 years \\ 
        Pro4 & Male & Surgeon& GI cancer and benign disease& 10-20 years\\ 
        Pro5 & Female &Surgeon& GI cancer and benign disease & 5-10 years  \\ 
        Pro6 & Female &Surgeon & Thoracic cancer&  10-20 years\\ 
        % \bottomrule
        \hline \hline
    \end{tabular}
    }
    \caption{Demographics of Provider Participants}
    \label{tab:provider-demographics}
\end{table}
\renewcommand{\arraystretch}{1.0}

\subsection{Participants}
After obtaining IRB approval, we used snowball sampling to recruit N=6 cancer healthcare providers from hospitals and clinics in the United States, as listed in Table \ref{tab:provider-demographics}. The providers include five cancer surgeons and one nurse practitioners (NPs) in hospitals or clinic offices. All participants have rich experience communicating with patients after cancer treatments.
We mainly recruit cancer surgeons for two reasons. Firstly, the patients who have gone through surgery have a more significant change in their health conditions and thus have a higher risk of having complications and being readmitted to the hospital, leading to a stronger need for patient-provider communication.
Secondly, although NPs can be involved in post-treatment communication, patients usually get directed to the doctors for actual clinical decisions and discussion, which corresponds to the focus of this study.

We also recruited N=8 cancer patients by posting recruitment information on social network sites and using snowball sampling. All patients have gone through cancer treatments, such as surgery, chemotherapy, or radiotherapy, and live in a single-home setting in the United States. Most participants live with family members to support them during and after their cancer treatment, while some have professional caregivers, dependents, or usually live alone. Some patients are also family members or caregivers of other cancer patients. We summarize their demographics and treatment backgrounds in Table \ref{tab:study1_patient}.

\renewcommand{\arraystretch}{1.3}
\begin{table}[b]
    \centering
    \resizebox{1\linewidth}{!}{
    \begin{tabular}{c|c|c|c|c|c|c}
    % \toprule
    \hline \hline
        \textbf{CP\#} & \textbf{Female} & \textbf{Cancer Type}&\textbf{Caregiver}& \textbf{Cancer treatment} & \textbf{Home setting} & \textbf{Diagnosis}\\
        \hline
        CP1& Male & Thyroid and lung cancer&Yes& Surgery, radiotherapy, chemotherapy& Live with family support & 2017\\ 
        CP2& Female & Spinal cancer&No& Surgery, radiotherapy& Live with dependents & 2016\\ 
        CP3& Female & Breast cancer &No& Surgery, radiotherapy, chemotherapy& Live with family support & 2020\\ 
        CP4& Female & Lung cancer &No& Surgery, radiotherapy, chemotherapy& Live with family support & 2021\\ 
        CP5& Female & Breast cancer &No& Radiotherapy, chemotherapy& Live with family support& 2021\\ 
 CP6& Female & Skin cancer &Yes& Surgery&Live alone& 2020\\
 CP7& Female & Skin cancer &No& Chemotherapy&Live with professional caregiver & 2022\\
  CP8& Female & Breast cancer &No&Chemotherapy, surgery, radiotherapy & Live with family support & 2022\\
         \hline \hline
    \end{tabular}
    }
    \caption{Demographics of Cancer Patient Participants. The column ``Caregiver'' represents whether the participant, despite being a patient, serves as a caregiver of another cancer patient.}
    \label{tab:study1_patient}
\end{table}
\renewcommand{\arraystretch}{1.0}

\subsection{Procedure}

We conducted semi-structured interviews with both groups about their communication experiences during or after patients' cancer treatments.  For cancer patients, we first asked them to share patient-provider communication examples, their communication needs and attitudes during and after cancer treatments. For healthcare providers, we asked them to share communication examples and focused on the crucial information and time points that they were looking for. After the contextual inquiry, we encouraged participants to share their ideas on how future AI technologies could support patient-provider communication. If participants were unfamiliar with AI, we shared examples like LLM-powered CA, data annotations, and risk prediction models to help them picture their expectations and concerns.

The patient interviews lasted 33 to 47 minutes (41 minutes on average). The interviews for providers lasted 27 to 45 minutes (33 minutes on average). The participants were compensated \$ 20 for their time. The interviews were recorded via Zoom and transcribed afterward. Two authors iteratively coded interview transcripts until a consensus was reached~\cite{corbin1990grounded}. Thematic analysis was then employed to analyze the data~\cite{braun2006using, glaser1968discovery}. We summarize our findings in Section \ref{4-findings}.
% The transcriptions are analyzed using thematic coding. Two researchers first independently coded the transcripts iterated the codes and then discussed the codes and themes until they reached an agreement. Then, they used the codes to finish coding the rest of the transcripts. 

\section{Findings}
\label{4-findings}
% sepsis: flow chart
% timing? timing

% figure 1: current practice, difficulty
% why not work

% figure 2: our proposal, how to work
% flow chart? communication topic?
Based on our interview results, we first provide an overview of current patient-provider communication practices in post-treatment cancer care, highlighting the communication timeline, needs, and existing technologies in Section \ref{results:overview}. Then, following our research questions, we summarize communication gaps between cancer patients and providers, the communication challenges within the cancer support network, and present AI potentials in filling the gaps and supporting collaborative cancer care in Section \ref{4-findings-what-and-when} and \ref{results:collaboration}. Lastly, we reflect on participants' conflicting attitudes towards leveraging AI for their patient-provider communication, revealing the sensitive nature of post-treatment cancer care in Section~\ref{sub:4-findings-sensitive}.

 % both groups mentioned their expectations for future technologies to support patient-provider communication and expressed excitement toward AI-powered systems for their communication after the illustration of a demo system. In the following sections, we present the overarching themes for the participants' expectations towards such AI-powered systems.

\subsection{Overview of Current Cancer Patient-Provider Communication Practices}
\label{results:overview}

\begin{figure}
    \centering
    \includegraphics[width=\linewidth]{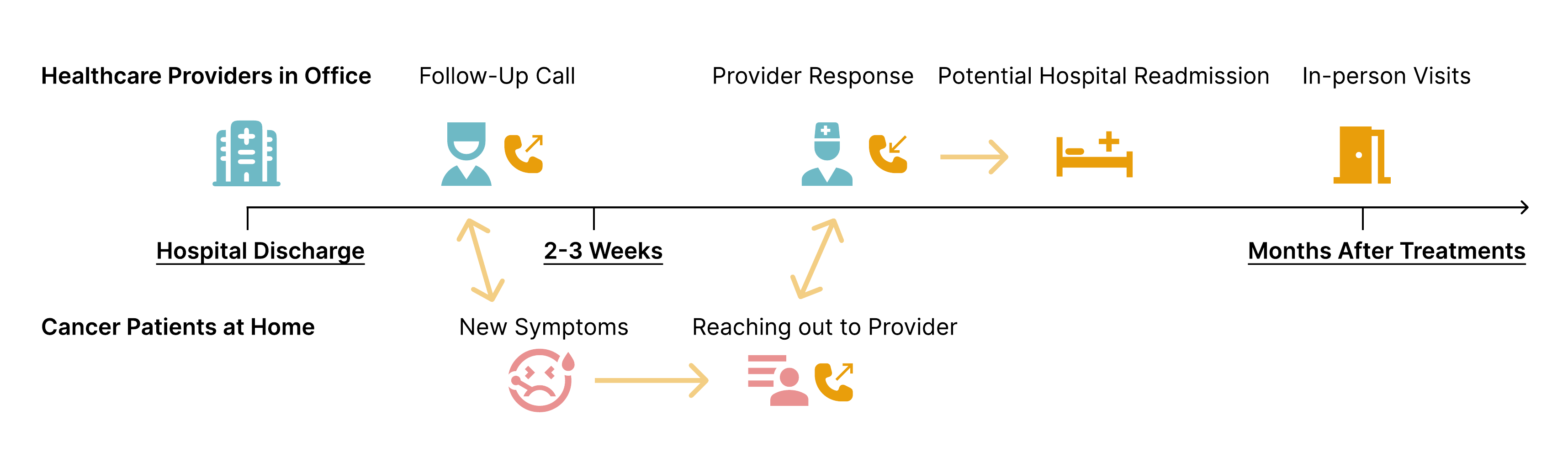}
    \caption{An example of Current Patient-Provider Communication Practice in Post-Treatment Cancer Care. After patients are discharged from the hospital, the healthcare provider may follow up via phone calls to check patients' conditions in two to three weeks. If patients have any discomfort, they can also reach out to providers. The key provider can respond remotely or readmit the patient for any abnormalities. Many patients have long-term follow-up visits to get lab tests.}
    \label{fig:communication-practice}
\end{figure}
\textbf{Procedures, Timelines, and Needs}\indent
Our participants mentioned different procedures and timelines for patient-provider communication after the patients are discharged from the hospital after their cancer treatments. The practices vary based on the type of treatments, the patient's visits and conditions, and the initiatives of patients and providers. As a typical follow-up procedure, some participants reported having regular in-person visits to the clinic. The healthcare provider checks on their situation from one to two months after cancer surgery (\Pro{4}) to yearly follow-up lasting over five years (\CP{2}). In this study, given the high readmission rates within 30 days from surgeries and the remote communication challenges, we mainly focus on patient-provider communication for cancer patients at home in a shorter period of time.

When patients are at home after treatments, multiple participants reported that the providers reach out to patients from the clinic to check on their conditions shortly after discharge, usually within the first two to three weeks. Some providers are motivated to call the patients more frequently or throughout a longer period (\Pro{6}, \CP{6}). The participants explain that within two to three weeks after surgery is usually when most problems arise, but the patients are still not familiar with the situation (\Pro{3, 4, 5}). For some patients, healthcare providers (especially NPs and therapists) would visit patients' homes to treat the patient or follow up (\Pro{1}, \CP{7}). Most of the time, the communication is intended for healthcare providers to check the overall condition of patients, or inform the patient of updated lab results (\Pro{3, 4}, \CP{1}).

The participants reported more situations when patients reach out to their providers to ask about questions or symptoms that they have or schedule appointments, and many of the phone calls are first navigated to a call center (\Pro{6}) or a nurse practitioner (\Pro{1, 2, 4, 5}) before the provider could answer. Some patients explained that they seldom communicate with their healthcare providers at home because they have regular in-person visits after the cancer treatments, and they did not have significant needs to communicate in addition to the visits (\CP{2, 7}). The patients mentioned various reasons for contacting the providers, mainly new symptoms, post-treatment complications, confusion about provider instructions or care-taking, and mental issues. When a health problem is discussed during the communication process, healthcare providers respond to patients in different ways. In serious cases, the provider schedules a clinic visit and readmits the patient to the hospital; in other cases, the provider informs the patient or calls the patient back to give instructions or suggestions. If it was not a call from the doctor, the response time might take several hours to days, or much longer. Therefore, a significant patient concern is the long wait before receiving treatments or responses, as we will discuss in Section \ref{results:timing}.\\

\noindent\textbf{Communication Technologies}\indent
Our participants reported mainly using phone calls for patient-provider communication at home, with over six mentioning using an online patient portal. 
Other technologies mentioned include video visits, emails, instant messaging apps with providers, and social media support groups for cancer patients.

% no participants report significant use of other novel applications, such as mHealth applications designed for cancer patients~\cite{jacobsMyPathInvestigatingBreast2018}

% When discussing the communication technologies that they are currently using, the participants discussed the advantages and limitations of these methods. 
Compared to phone calls, multiple surgeon participants favored using the patient portal to communicate with patients asynchronously for its accuracy, efficiency, and timing benefits. \Pro{5} mentioned that she recommends the patient portal to patients because messages are in patients' own words, and the information gets recorded into the system. Three cancer providers also suggested that asynchronous tools such as patient portals fit into their busy schedules after a long time in the operating room (OR).

\subsection{Filling in the Gap: When and What to Communicate}
% Potential Title: \\
% Filling in Communication Gaps between the Patient and the Provider with AI \\
% Determining the Appropriate Timing and Content for Communication with Doctors: AI in the Mix
\label{sub:results:appropriate-timing-content}
\label{4-findings-what-and-when}

We recognized distinct expectations in patient-provider communication in post-treatment cancer care. In this section, we will discuss the main gaps between patients and providers, the emotional needs, and AI potential: ``\textbf{When} and \textbf{what} should I talk to my provider, and can AI help me?''
% We summarize these challenges into the following three aspects, focusing on the information communicated, the failure in communication, and the additional emotional care needs. 
% what and when should I talk to my doctor? and, can AI help me?

\subsubsection{Knowledge Gap: What to Communicate and What Not to}
\label{results:knowledge}
From our thematic analysis, we identify the knowledge gap between cancer patients and providers, leading to their communication challenges in exchanging information. The providers have rich experience treating cancer patients, while many cancer patients do not know much about cancer until their diagnosis or even treatments. As a result, patients and providers find it difficult to communicate with each other effectively. We recognized a deficiency in the current strategic approach CP and Pro used in practice, underscoring the necessity for advanced technologies like AI to intervene and provide support.\\

\noindent\textbf{Disparities in Identifying Crucial Information}\indent
\label{4-findings:knowledge:crucial}
During our interviews with healthcare providers, they were consistently clear about which pieces of information were crucial. Most providers first look for information about the patient's overall condition, such as eating, drinking, lab results, fever, or specific symptoms related to the operation (\Pro{2, 3, 4, 5}). 
% For example, providers of colorectal cancer patients would seek information about drain output, bowel movements, and hydration (\Pro{2}, \Pro{3}). 
Since providers have rich experience with patients, including readmissions and complications, they directly look for key information during communication. Given that providers have a clear logic in mind, they typically expect patients to also efficiently provide crucial information as needed. However, patients with less professional information may struggle to meet these expectations. Interestingly, \Pro{4} categorized three groups of patients:
\begin{quote}
    \pquote{4}{One is people who never call for anything. They could be dying, (but) they don't call, right? ... And then the middle ones are more reasonable to have some real concerns... And then the third type is they call for every little thing, doesn't matter how small or big, and show me every single picture. It's so painful.}
\end{quote}
The challenge of not being able to efficiently obtain the appropriate amount of essential information from patients places a significant burden on providers. Meanwhile, it is also frustrating for patients not knowing what to ask and being unable to fulfill doctors' expectations. As \CP{6} said, \textit{``I just wished I knew more.''}, patients do not know what questions they should ask the providers.

% \textit{I just wished I knew more. ... Questions would come up, ... and then you'd forget, we could go into the doctor's office, and then you'd forget all the questions that you had. (if possible, find a better quote that emphasize not knowing instead of forgetting)} (\CP{6}). 
The reason behind this lack of knowledge of symptoms is intriguing, especially when considering the availability of supportive materials like post-surgery care instructions. 
In discussions with healthcare providers regarding their approaches to diagnosis and treatment, we discovered that while they adhered to a basic structure, the specific pathways were often customized and dynamic based on patients' answers.
In the case of abnormal symptoms, providers would investigate detailed and follow-up information on various aspects. For example, \Pro{3}, who was particularly concerned about ileostomy output, would make a series of informed and professional decisions based on personalized situations in identifying the issues and proposing solutions:
\begin{quote}
    \pquote{3}{...if it's too high, if it's too low, [any] risk for constipation, dehydration, if no, is it early obstruction, or do they need fluids?}
\end{quote}

Hence, other than a fixed instructions or checklist, there was a need for a more customized, adaptive, and responsive process to enable patients to better discern crucial information, particularly concerning symptoms, in varying situations.\\

\noindent\textbf{Challenges in Literacy and Understanding}\indent
In addition to the varying levels of understanding regarding the importance of information, there is also a knowledge gap in literacy between patients and providers, especially considering the special expressions in medical settings.
The initial challenge lies in the language barrier when communicating with non-English speakers, significantly impacting patients' understanding and adherence to instructions. For example, \Pro{4} shared the difficulty communicating with a Spanish-speaking patient. Even though the healthcare provider team tried translating the documents and talking to an English-speaking family member multiple times, the patient did not follow the instructions and ended up being readmitted to the hospital due to a minor issue that could have been avoided. 

Another barrier for patients is their limited literacy compared to the providers, particularly in understanding highly technical and professional expressions. P6, reflecting on their over 20 years of medical education and training, realized that physicians were using 'a language' that they often used with other physicians to also communicate with patients who did not have that training and experience'. More specifically, we noticed significant disparities in the language providers and patients use to describe patient conditions during our interviews. For example, a patient would say \epquote{2}{I'm in a lot of pain}, while a provider would have precise descriptions like \pquote{3}{worsening pain} or \pquote{4}{abdominal pain}. When patients mention their surgery in the part of the body like the breast or lung, providers use much more professional keywords like \textit{``rectal'' ``lower resection'' ``divert those with an ostomy''} or \textit{``anastomosis''}. 

% For instance, talking about a patient who had a surgery, 

% \begin{quote}
%     inflammation, you know, versus like their rectal cancer patients,...So especially if it's like a lower resection, usually we will divert those with an ostomy to protect the anastomosis...
% \end{quote}
Realizing this issue, some providers started to give patients instructions with highlights, reviewed the contents with patients, and explained key concepts(\Pro{3, 4, 6}). However, this would still take extra time and effort. The barriers in current strategies, such as file translation and one-on-one explanations of concepts used by providers to bridge the literacy gap, highlight the necessity for a more efficient and intelligent system to adeptly address the linguistic aspects of diverse materials.\\

\noindent\textbf{Inaccurate Expectations of the Cancer Journey}\indent
Other than the gap related to symptoms and expressions, patients' and providers' different understandings of the overall cancer journey also affect their communication in various ways.
Realizing their limited knowledge about cancer, many patients learned a lot about their cancer, treatments, and survival outcomes after their diagnosis through various methods. Their learning about cancer from healthcare providers, their social network as well as online resources not only helped them understand the treatments and instructions but also lifted their spirits toward their cancer treatments considerably (\CP{3, 4}). For instance, after \CP{4} learned that she had a good chance of survival in the early stage, she was encouraged to move on with the treatments, adding \epquote{4}{now I believe I'm stronger}.

Correspondingly, all providers shared ways of educating patients about cancer, taking different strategies. For example, \Pro{2} shared that he would \textit{``educate them at every step of the surgery''}, from explaining what the surgery is for to what patients should expect after the surgery.  Patients also shared that they find the provider's suggestion on community support, books, and papers helpful.
However, this required excessive effort and was not a common practice. Some patients still searched online and believed in fake information. as a result, advanced solutions are needed.\\
% However, this required excessive effort and was not a common practice. (other reasons why the current strategic is not enough) some patients still searched online and believed in fake information. as a result, advanced solutions are needed.\\

% Similarly, looking back on their cancer journey, the patients advocated for more awareness of their symptoms among the general public and cancer patients. For example, \CP{4} said,
% \begin{quote}
%     \textit{So I feel there should be an awareness to cancer patients on the line the same time, so if they actually start coughing, and it's really severe, ... and they should go for an checkup immediately in order to him to avoid the spread of the of the cancer you get. }
% \end{quote}

% \CP{3}: And also, I learned a lot from the internet, and by reading books and articles and talking to people who have a spirit experience with cancer, because, you know, at that point that I was so scared, I was ready to read any everything and know, okay, I will definitely be able to make it.  

% \CP{4}: So when a friend of mine told me, she tried encouraging me and said, I shouldn't be scared that it's at its early stage, it can be caught. So I bought the idea. I tried to be confident at that time. So I, she actually told me up suffered from leukemia and all those jobs, but actually, she's still living through now. So I took that as an encouragement. . I'm going to leave. So at first I was scared, but now I believe I'm stronger. I don't have such fears anymore. 

\subsubsection{Timing Gap: When to Communicate and When to Get Reply}

As we mentioned in Section \ref{4-findings:knowledge:crucial}, providers summarized two extremes of participants who either contact too much or too little. This encompasses not only the content but also the frequency. To understand the causes that lie behind these communication challenges, we delve deeper into their remote communication to reveal the following timing gaps.\\

\label{results:timing}
\noindent\textbf{Patients' Need for Responsiveness versus Providers' Busy Workflow}\indent
We first find there is a group of patient participants who actively reach out to their providers. Talking about their experiences communicating with healthcare providers about their concerns, multiple patients emphasized the importance of quick and in-time responses from their providers. Multiple causes result in the need for responsiveness during patient-provider communication for post-treatment cancer care. The patients' confusion, limited knowledge, and worry about the high risk of cancer have led to their urgent need for providers' response.
\begin{quote}
    \epquote{3}{That was actually a challenge for me at the moment ... I don't actually want to go through any risks at the moment because you don't know if maybe a second of anything could lead to something.} 
\end{quote}
Some providers managed to respond to their cancer patients promptly, while many others did not, causing the patients' frustration and possibly higher risk to the patient's health. \Pro{1} revealed that in order to be responsive to her patients, she gave the patients' families her personal phone number and visited patients \pquote{1}{in the middle of the night}. On the contrary, \CP{4} and her family were unsatisfied with the provider's responsiveness. Thus, the family member argued that the provider was \textit{``incompetent''}, \epquote{4}{What if something had happened to me?}. 
% \epquote{4}{... he didn't respond immediately. (... my brother) was like telling me and that's incompetence in his own as a health care provider. He's not having time for his patients is really, really bad. What if something had happened to me or stuff like that?}

% \CP{4}: Like I said, sometimes I need help dealing with my trauma and anxiety. So me calling him okay, there was a time I recalled when I was feeling so on easy. And I knew I knew I knew something wasn't right, right about the the way I was feeling. So I tried. I tried to reach him. And he didn't respond immediately. I thank God that day, my my brother came home quite early. And I told him about it. He was like telling me and that's incompetence in his own as an health care provider. He not having time for his patients is really, really bad. What if something had happened to me or stuff like that? So I was like, you know, I was like, you know, I was, I'm not the only patient he has been. He has been listening to him. So he should try and come down. 
However, as we see from our interviews, providers did not intentionally delay the response but struggled to reply. Three provider participants report that they spend 70\% to 80\% of their time or more in the operating room (OR), and thus, their time to respond to patients is extremely limited.
\begin{quote}
    \pquote{4}{hugging the patients takes a lot of time and it's very unpredictable. And it's a huge detriment to my life. Because, ... after 12 hours a surgery, I still have to call people back for two, three hours. ... I don't want to do that. }
\end{quote}
% need a summary
Observing this tension, we recognize that current strategies were insufficient to strike a balance between patients' urgent needs and providers' limited time. This calls for a future system to offer assistance to patients while aiding providers in managing their schedules.\\

\noindent
% \textbf{Urgent But Unsuccessful Communication}
\textbf{Communication Delays in Addressing Urgent Signals}\indent
Given the knowledge gap in cancer, patients also fail to communicate with providers in a timely manner, leading to hospital readmissions and health risks. Two provider participants specifically illustrated outpatient visits that could have been avoided because the patient did not reach out to them earlier (P3, P4). For example, one colorectal surgeon mentioned one patient experiencing dehydration but did not contact her early, \pquote{3}{(he could have) started antibiotics earlier. So he didn't need to be admitted to the hospital}.

Admittedly, some cancer patients do not actively take the initiative to communicate with their providers early, their unawareness aside. During our interviews, multiple patients shared that, after failing to get an in-time response from providers, they felt less comfortable reaching out to their providers and thus sought support from family and friends instead of contacting providers (\CP{5}). Therefore, both groups expect future technologies to help patients be more aware of the symptoms and thus communicate more. 

\subsubsection{Patients' Fear: A Call for Attention and Reassurance} % emotional gap

Built upon the knowledge gap and timing gap between cancer patients and their providers, the emotional gap between fearful cancer patients and calm providers also introduces communication challenges. Six of the patient participants and two of the provider participants specifically mentioned that patients were scared when they were diagnosed with cancer, largely due to their limited knowledge of cancer and treatments. \epquote{3}{"I felt the world was actually over for me, like I had no choice anymore, I was I even had a thought of dying."} . Some patients have confusion when they start to develop new symptoms, feeling \epquote{4}{I've never had that before. }.
% \textit{That was persistent. I've never had that before. So I was very, very confused. So when I started coughing, and I started losing appetites, and sometimes shortness of breath.}(\CP{4}). 
The participants expressed a sense of hopelessness because they once believed that no available cancer treatment existed, leading them to perceive that death in the near future was inevitable. Because of the mental burden, two of our patient participants have experienced mental health issues such as depression and anxiety, forcing them to seek professional support from therapists (\CP{3} and \CP{4}).

Acknowledging this issue, both parties emphasized patients' need to get reassurance from their providers about their health conditions and treatments. Some providers have taken action to support patients by explaining the options and risks. \pquote{4}{(if not) they're just gonna call me right back and ask me am I gonna die or something like that.}
Moreover, some providers reassured family members that they could support the cancer patient, and the family became \pquote{1}{very proud of themselves}.
% \textit{So I was able to reassure people that they would learn how to do something, and be very proud of themselves afterwards. } (P1). 
Effective patient-provider communication on emotion can expand the patients' knowledge and lift up the patients' attitudes. However, we noticed in the interviews that this reassurance is often absent due to patients' limited opportunities to communicate with providers, as we found in \ref{results:timing}. Many patients expect providers to pay more attention to them and express care and emotional support. Some patients are disappointed with their providers when they feel that the providers are just going through designed procedures with them (\CP{3}), saying \epquote{6}{they kind of don't care}. In our interviews, we did realize that only two providers reflected on their efforts addressing patients' mental needs, while the majority of providers focused on patients' physical health rather than emotional concern. Despite the urgent emotional needs of patients, they are typically not the highest priority for providers, especially given their busy schedules.
\subsubsection{AI Potential: Filling in the Communication Gap with Knowledge and Interaction}\hfill

%In our interview, we briefly introduced an LLM-powered system as an example of AI to support patient-provider communication and invited our participants to share their opinions. From the knowledge perspective, participants mainly commented on the LLM-powered CA to interactively talk with patients, approving the potential promotion of patients' learning.

% awareness and reflection
% ljc: awareness of the symptoms
\noindent\textbf{AI Interactive Agents for Knowledge Gap}\indent
Firstly, both providers and patients expressed the need for an AI assistant to check in with patients regularly to improve patients' awareness and reflections on their new symptoms. This could be particularly useful in abnormal cases so that they can take action on it immediately. For example, \CP{1} thought having this kind of LLM-powered CA could help him notice his calcium drop issue earlier. 
\begin{quote}
    \epquote{1}{But yeah, before I knew that the calcium was that I was going downhill without it. If it asked questions ... it would have noticed the symptoms. ... I think it will recognize the patterns in symptoms ... to suggest contacting the physician...}
\end{quote}
Similarly, \CP{6} also agreed that such a system could have helped her reflect on her fever symptoms that could be related to cancer complications. Thus, participants look to future technologies to support their awareness and thinking about symptoms interactively.

%customized and follow-up
% \textbf{Follow-up Questions for Important Symptoms}
% ljc: follow-up questions for important symptoms\\
% During our interviews, we observed that healthcare providers possess a well-defined logic regarding the specific follow-up questions to ask in various situations. Many of these situations are dynamic, contingent on the patients' responses to symptoms. Unfortunately, patients often lack this understanding, hindering their ability to uncover the full scope of their unique issues effectively. 
In our interviews, we noted the significance of providers dynamically posing follow-up questions based on patients' responses. Recognizing the absence of this knowledge among patients, some providers expressed agreement that it would be beneficial if the check-in system could also delve deeper and include follow-up questions. For example, \Pro{6} mentioned patients may \textit{``downplay''} their symptoms, and thus AI could ask more questions to probe deeper, and patients would say \textit{``well, actually, ...''} followed by their actual conditions.
% \begin{quote}
%     \pquote{6}{... it's nice that it does the follow up question because a lot of people will be like, `oh, yeah, I'm fine.' that's why I call them because I probe deeper... then they're like, `well, actually, ...'  And so actually, [if] you rely solely ... on patient reported [symptoms] they're gonna downplay some of their symptoms sometimes ...}
% \end{quote}

% \begin{quote}
%     \pquote{6}{P6: Because the the follow up, it's nice that it does the follow up question because a lot of people will be like, oh, yeah, I'm fine. And then and then when you probe deeper, that's why I call them because I probe deeper. I'm like, Yeah, but did you go for a walk today? Are you doing a deep breathing coughing? And then they're like, Wow, actually. And so actually, you rely solely I think on Patient Reported they're gonna downplay some of their symptoms sometimes, as opposed to, if you actually follow up and say, Hey, ya know, I really do want to hear if there's something that's not right, because that's what prevents admissions readmissions is actually drilling down and finding those things that are really like that.}
% \end{quote}

% ljc: explaining in a plain language
Our providers agree that the LLM-powered CA could support their work communicating with patients by explaining cancer treatments and instructions and providing related information. Both \Pro{2} and \Pro{4} suggested that an AI system might be able to tell common symptoms and answer simple questions. 
\begin{quote}
    \pquote{2}{The AI system could prepare the patients properly for the procedure, ... Tell them what they're going to expect from the first week, ... What they're going to expect post-op time. ...  It could teach them and guide them every step of the procedure. } 
\end{quote}

P6 added that, given the vast amount of online health information, AI could filter out unreliable information for patients. Patient participants are also interested in such AI to answer simple questions about their confusion and have attempted to check that information online(\CP{6}). \\

% \epquote{6}{Sometimes, when I need an answer to stuff or when I feel I don't understand this particular thing... I tried to check it on the internet or using the AI chatbox. }

% \pquote{2}{And some of the common questions can be answered by the AI system ... Some things are normal, or every pre-op patient has to go through certain kind of testing and pre-op preparation, to go for a big fuzzy that can be common to all the patients. }

% As \CP{4} discussed, \pquote{4}{Some might be suffering from it, but they might have seen the symptoms, right? Lack of awareness, so they don't know until it gets very worse, which is so bad. }

% \CP{4}: Some don't even know have any knowledge about the symptoms. Some might be suffering from it, but they might be seen the symptoms, what? Lack of awareness, so they don't know until it gets very worse, which is so bad. So I feel there should be an awareness and to cancer patients on the line the same time, so if they actually start coughing, and it's really severe, they should know something is actually wrong with them, and they should go for an checkup immediately in order to him to avoid the spread of the of the cancer you get. 

% \subsubsection{AI Potential: An Efficient and Transparent Workflow}

\noindent\textbf{An Efficient and Transparent Workflow}\indent
Firstly, providers look forward to having an AI-powered CA to complete some tasks for them, including having conversations, probing for details, and providing transcripts and highlights for review (\Pro{6}). 

% \pquote{6}{I actually think that's really neat. Because you could actually have that sort of conversation with them. It could request additional details that ... you could get like a little transcript of it saying, Hey, here's the issues we've highlighted that this patient wants to report to you.}

Secondly, some providers commented on the different levels of risk and reckoned that such systems should be able to differentiate changes and abnormalities. \pquote{2}{And anything which is abnormal in the sense that AI system should contact the office of the doctor immediately... } P6 also suggested using \textit{``buzzwords''} to escalate some findings more than others.

Providers provided different interaction designs for such systems to provide reminders and alerts. Three participants discussed sending these risky cases as alerts instead of requiring someone to monitor a dashboard, \pquote{5}{we're not logging into the dashboard all the time}.
% \begin{quote}
%     \pquote{5}{... we're not logging into the dashboard all the time. So, are there certain like alerts that are able to be presented to us...because if the system requires that we log in to look at the dashboard, I can guarantee that that's probably not going to happen...}
% \end{quote}
Providers added that alerts could promote in-time communication, \pquote{6}{We don't want to lose that window of opportunity} The alerts would help providers make the most out of the system without extra burden.
% \pquote{6}{We don't want to lose that window of opportunity to fix the patient properly.} 

Lastly, our provider participants believe that asynchronousity in an AI-powered system to support communication could promote efficiency in their workflow. For instance, NPs will be able to respond new requests after the prior task is completed, \pquote{4}{[it will] give her some control in her workflow}. Considering that such a system could respond to patients and shift communication from synchronous to asynchronous, the providers could have better control of their workflow.\\

% \pquote{4}{... she can finish task A and then, just check in and be like, Okay, well, let me check who called and who was concerned. Let me call these people back proactively, you know, gives her some control in her workflow.} 

% will need to connect the two parts
\noindent\textbf{Concerns on Overloading and Utility}\indent
Regarding AI's potential to fill in the timing gap by engaging in conversation, a significant concern for healthcare providers is the additional information introduced with such systems. Some participants are concerned that such systems urge human clinicians to provide quick responses, \pquote{1}{And I guess one of my concerns would be that this kind of intervention ... really does require action. And that puts more pressure on a clinician.}, while some others added that they do not hope to get information that is not crucial, \pquote{4}{I don't want to report on how bad my patient was every single day. ... That's too much.} 

Similarly, although participants agree that frequent checks by AI may improve patients' awareness and fill in the timing gap, such systems should also avoid communicating with patients too frequently so that it gets annoying, else\pquote{1}{I would just dismiss [AI]... be angry at [AI].}

Regarding the responsiveness of such systems, our participants question whether the systems reach providers for patients instead of merely responding but not solving problems. \epquote{6}{here's the key is there has to be the linkage has got to be there. If you get all the information, and then it's forwarded to the healthcare provider and nothing happens, then I think that the person will lose confidence that the system is responsive. } Similarly, providers brought up the example of calling 911 in automatic phone calls, \pquote{4}{it doesn't help anybody. And it's very frustrating.}. Instead, they suggested that feedback that responds to patients' needs, \pquote{4}{I think you should ... expect a call from Dr. XX's office in 20 minutes to discuss your symptoms and ... talk about some rehydration strategies...}, which is specific and respond to patients' questions.\\

\noindent\textbf{Providing Mental Support}\indent
Four providers agreed that the AI-powered CA will be able to support patients mentally by expressing reassurance, attention, and care. Having a conversational agent able to perform natural conversation may comfort the patients, \pquote{1}{it is the feeling that something cares.} Participants added that this might be particularly comforting for patients staying at home by themselves by eliminating the loneliness, which could also be applied to other patient populations, since \pquote{1}{any attention is better than no attention}.

Providers also shared the potential of reducing patients' fearfulness with an AI-powered CA to check in regularly. \pquote{6}{... There's also a lot of fearfulness of medicine ...} Thus, patients may get accustomed to talking about their treatments and symptoms by interacting with an AI frequently and thus feel less fearful of the illness.

% P6: . There's also a lot of fearfulness of medicine, and there's some people who are fearful of technology, sort of like age aside.

% But ... I think half of the potential participants would respond, it is the feeling that some something cares. You know, and that there is concern whether it's human or not, but and the fact that the, I hope that the patient knows that, that that information will be transferred back to the clinician.

% P1: But I picture elderly living at home and I know from from the patients that are from the people that we've seen now. Non cancer, but still very serious disease states that any attention is better than no attention.

% P6: . There's also a lot of fearfulness of medicine, and there's some people who are fearful of technology, sort of like age aside.
% \subsection{Communication Challenges In Collaborative Cancer Care}

% \subsection{Post-Treatment Care Is a Team Sport and AI Needs to Be a Team Player}
% \textbf{Post-Treatment Care: A Team Sport that Needs a Manager}\\
\subsection{Collaborative Cancer Care: The Team Sport in Need of a Manager}
\label{results:collaboration}
% Interestingly, we realize that communication in post-treatment cancer care not only exists between the two distinct patient and provider groups but also involves the close collaboration of circles of stakeholders. 
While patients and the primary cancer care provider (e.g., surgeons for surgery treatments, oncologists for chemotherapy treatments) are centers of communication after cancer treatments, we uncover through our study that post-treatment cancer care involves the close collaboration of circles of stakeholders. 
As communication challenges rise within the support network, we conclude that post-treatment cancer care is a team sport, and AI needs to be a team player.

% post-treatment cancer care is a team sport and AI needs to be a team player
\subsubsection{Network of Healthcare Providers}

Participants shared examples of collaboration between healthcare providers. Surgeon participants commented that having nurse practitioners or assistants to contact patients for them has lifted their workload. Both groups of participants mentioned having more than one healthcare provider throughout the whole cancer treatment process. There were not only surgeons and nurse practitioners but also primary care providers (PCP), oncologists, specialists related to the specific type of cancer (e.g., neurologist for a spinal cancer patient), and sometimes ER doctors and mental therapists. However, patients do not always reach the providers who are the most familiar with their treatments, causing risks and challenges.

Even though the surgeons are the ones most familiar with patients' treatments and potential issues in post-treatment cancer care, healthcare providers mentioned not knowing about their patients' conditions until they were informed by doctors from Emergency Rooms (ER).  
\begin{quote}
    \pquote{2}{We have one patient ... went to the outside hospital... and the ER doctor called us and told me to get proper treatment for them. So we called them back and told them what to do. And the patient recovered within two days and then went back home.} 
\end{quote}

In those cases, the ER doctors lacked a tool to promptly acquire the treatment-related issues of the patients, so close communication with the corresponding surgeon should be emphasized. A future personal assistant is crucial to assume the role of managing all relationships and information within the patient's healthcare provider team.

\subsubsection{Support Network around Patients and Providers}

% Communicate with family and friends -> 1. doctors also communicate with family, more involved
% 2. sometimes unprofessional, issues

% Almost all patient participants and also some provider participants reported an expansion of their support group after their cancer diagnosis. Over half of the patients mentioned that they lived alone before they were diagnosed, but their families moved in when they were receiving the treatments, both for physical caregiving and mental support. \pquote{}{} % afraid of being alone, have brother as company 
% Some healthcare providers managed to help expand the patients' support network by reaching out to more family members or introducing patient community engagement. \pquote{1}{... in the end, she died. But he became her support network. And was responsive to it. He had just lost touch with her and didn't know...}

Other than the professionals, some family members and friends also played the role of healthcare providers to some extent, offering additional assistance when primary providers are unavailable but also introducing risks into the caregiving process. For example, one participant reached out to her family about her post-treatment concerns when the provider was not responsive, and her family found friends who are healthcare professionals to help her decide on treatments. However, since these friends are not communicating with their own cancer care provider, suggestions from the support network may not be suitable for the patient.
\begin{quote}
    \epquote{6}{I'm lucky, we're lucky that we have friends who are doctors. And so we can kind of have a backdoor. But those doctors ... they're not going to diagnose somebody else's patient. You know, they can give some general information. }. 
\end{quote}

% Yet participants also mentioned that providers are not always available when they need this new information, However, as discussed in related literature, we found knowledge from other resources introduced confusion and health risks to patients. Thus, participants call for future technologies to improve the reliability and accessibility of cancer-related health information.

Other than stakeholders around patients, providers also have a support network around them. When \Pro{1} visited her patients in the middle of the night for emergencies, her husband drove her to the site and waited, \pquote{1}{one of our patients... called him Mr. Nightingale}. Although the collaboration may help with patients' immediate needs, the patients' inevitable communication with the providers' support group results in extra burdens and may lead to misunderstandings or health risks. 
Consequently, the future system also needs to function as a 'relationships manager,' acknowledging and overseeing the social support network around patients and providers within the caregiving process.

 % It's these, my my husband actually would go out sometimes in different parts of the city would go out with me, if it was middle of the night, and one of our patients, sent him a note and called him Mr. Nightingale. Because he, he would just drive me and sit and wait in the car while I was doing what I needed to do, and that might be a procedure of some kind, I mean, a simple thing, whether it's, you know, whether it's oxygen or something of that nature, but But you did, we really did. respond as quickly as possible.

\subsubsection{Resource and Accessibility Challenges}
\label{results:resource}
While these communication challenges could be generalized to many patient populations, we also realize communication practices that are significantly challenging for cancer patients, and especially post-treatment cancer patients.\\

\noindent\textbf{Travel Inconvenience}\indent
More than one-third of patient participants specifically expressed their concern about the difficulty of traveling when communicating with their providers, especially for them as cancer patients and when some visits are not necessary. Traveling is particularly challenging for cancer patients and their caregivers because they are likely to be seriously ill, and traveling could lead to higher risks to their health. Both groups call for remote patient-provider communication for not only convenience but also health considerations.
\begin{quote}
    \epquote{6}{... they're a lot of times they're bedridden, and to take a cancer patient to the ER, like my sister, you know, ended up being, I think the cause of her death.}
\end{quote}

The challenge to travel becomes more significant due to the rarity of cancer. \Pro{1} shared an example of a cancer patient who \textit{``lost half of his body practically'' }due to massive surgeries, yet still needed to travel six to seven hours to get to his provider because there were no available providers for his cancer nearby.
% \textit{P1: And this was a gentleman who had lost half of his body practically, and was on heavy doses of pain medication. So he would call and say they won't refill it, they won't refill my oxycontin or they're only going to refill it for 30 days and I live 30 miles away from from a pharmacy. }
% P1: one of our patients who was a gentleman who lived far out in West Virginia. So he would call and say they won't refill it, they won't refill my oxycontin or they're only going to refill it for 30 days and I live 30 miles away from from a pharmacy. And they would travel for six or seven hours, never complaining
Therefore, participants agree that the availability of effective remote communication methods could greatly lift their healthcare burdens, contribute to their survival, and potentially expand the availability of cancer treatment for more cancer patients in areas with fewer medical resources.\\

\noindent\textbf{Medical Resources Inadequately Tailored to Cancer}\indent
Both patients and providers also expressed their frustration in communication due to limited medical resources for cancer patients, including the availability of medication and the capacity of hospitals and clinics. \Pro{1} described a case where she had to negotiate with multiple stakeholders, including the insurance company and the pharmacy, to get a patient's pain medication. Thus, the participant felt the communication was too taxing. \pquote{1}{we've seen the consequences of inappropriate use of narcotics and opioids, but the cancer patient that requires it often tends to get lost that these days ... And with someone, you know, like this gentleman, or many of my other patients, ... they require opioids. } 
In addition, \CP{1} mentioned his frustration of having to wait for a long time at the hospital to get tests, corresponding to P5's comments that the hospital is only able to take a limited number of patients one day. The participants suggested that, since many cancer patients are seriously ill and suffering from the disease, they should be considered with priority in these cases.

\subsubsection{AI Potential: Supporting Cancer Care Collaboration}

Given the complexity of the support groups in cancer care, both groups look forward to future AI technologies to support their collaboration. To help patients navigate the healthcare provider network, \CP{2} expressed the wish that there is an AI to route her to the provider responsible for the current issue, \textit{``I just want them to send it to the right doctor'' }(\CP{2}). AI's capability to analyze patient symptoms and matching with provider expertise or instructions could support patient navigation work and possibly reduce the workload of patient navigators or practitioners.

% \begin{quote}
%     PP2: So I want them to know to ask the right questions, you know, and to send it to the right doctor, right,
% \end{quote}

For the communication and collaboration between related healthcare providers, Patients also suggested that their medical records, confusion, and symptom updates could be managed by an AI assistant, and the information could be shared across the healthcare network, possibly with efficient AI documentation (\CP{6}).
% \epquote{6}{... it'd be really nice to have, like a robot be your primary care provider where ... he has got all your medical records and all your history and all your data, and the doctors, any doctor can access it. }
Similarly, some patients reckoned that AI could be the personal medical assistant not only to keep track of in-person appointments but decide on scheduling online meetings with multiple related healthcare providers together. 

% \epquote{6}{PP6: And I it seems like a perhaps a robot or an AI program could filter that stuff out, so that the doctor could come online? And and, you know, just spend a few minutes and could see a lot more patients that way. }

\begin{quote}
    \epquote{6}{But I was thinking, you know, can't we just do a little zoom call? ... The three of us talk.... it would be so cool to have, maybe a robot ... that's managing somebody's care, and keeping everybody informed of what's going on with that person.}
\end{quote}
In addition, providers thought there should be considerations on who receives the patients' information first, where NPs or patient navigators may be prioritized. For example, \CP{1} proposed that \textit{``someone that genuinely knows the patient''} should be contacted first.

% \pquote{1}{So I think someone that genuinely knows the patient (should get to it), I mean, I think it's helpful if it's someone who knows the patient.}

% P1: So I think someone that genuinely knows the patient (should get to it), I mean, I think it's helpful if it's someone who knows the patient. And certainly the nurse navigators, or the nurse practitioners or the PAs know, the patient's better than the than the surgeon. Usually,

% \textbf{AI Potential: Promoting Convenience and Coverage}

Although the participants feel it is a pity that they don't have a good solution for resource limitations, it is possible that we can address some of the communication issues and thus put the resources to better use. Some patients appreciate remote communication technologies to save their efforts from travel. The provider participants also favored having such AI systems to actively reach out to more patients since they do not have the time to call patients, but some patients remain passive in their communication. For example, \Pro{3} thought AI could \textit{``routinely follow up with everyone'',} which the providers cannot complete all the phone calls.

% \pquote{3}{I think it's nice to routinely follow up with everyone because sometimes we just get the patients that call. And not every patient with complications calls...}. 

% P3: I think it's nice to routinely follow up with everyone because sometimes we just get the patients that call. And not every patient with complications calls.... Because some people don't talk to us even though they're having issues.

% \subsection{An Accurate and Customized AI Is Not Necessarily a Trustworthy and Useful AI}
\subsection{Sensitive Nature of Cancer Care: Accuracy and Customization in AI Do Not Guarantee Trustworthiness and Utility}

\label{sub:4-findings-sensitive}

Apart from shared expectations between cancer patients and their healthcare providers, we also discover contradicting expectations and concerns regarding future AI technologies to support patient-provider communication. In this work, we draw upon these conflicts to reflect critically on our design implications for future intelligent systems.

\subsubsection{Conflicts between Customization, Personalization, and User Privacy}
In addition to the communication challenges shared among different patients, our participants also respond to the call for customization and personalization. For example, \Pro{1} reflected that the follow-up questions for children could be very different from those for older adults, \pquote{1}{there are reasons to use other scales}, while \CP{6} suggested that the questions for her as a cardiac patient should be different from others who are not. Thus, both groups showed interest in future AI-powered systems customized to the patient's health and treatment history when supporting communication.

% P1: And, you know, there are other reasons if it's a child, if, you know, there there other reasons to use other scales, but I think that's generally the one we use. And that's an appropriate question. 

To begin with, providers introduced a set of check-up questions that could be customized for post-operation patients, including overall feeling, eating, and drinking, as we summarized in Section \ref{results:overview}(P2). Some providers brought up the specific example of using standard scales such as pain scales in questions for LLMs to talk with patients. 
% \pquote{1}{I mean, pretty much everyone uses that that pain scale. And patients respond well to it and no matter what educational level... And I think we all sort of use that same standard.} 
They supported incorporating such protocols and standards designed by providers into an LLM-powered system to follow up on patients.

Secondly, they also emphasized the differences between patients, such as patients who underwent different surgeries. Thus, providers expressed their excitement about future AI systems that can help differentiate between patients and thus identify different risks. For example, \CP{6} suggested the follow-up questions for her, as a cancer patient with heart disease, should focus more on her discomfort in the heart, whereas her sister may be asked more about pain. Similarly, \CP{2} expected AI to know about her medical history.

% PP6: And I would assume that, that it would be programmed, the program would know what kind of questions it would be targeted for the kinds of complications that particular cancer patient might be experiencing? Or could experience. I know, after my heart surgery, you know, they would ask me questions about, you know, do I have my ankle swelling? You know, do I have, you know, do I feel like my heart's fluttering, things like that, specifically? Asking about what's going on? You know, for me, so I would think, if it's a very specific set of questions for a person, I'm trying to think like, you know, for my sister, you know, pain levels would be very important to ask, and, you know, things like swelling and any other symptoms, dizziness, things like that.
\begin{quote}
    \epquote{2}{I want them to know why. You know, her neck pain is probably from the arthritis from her surgery, cancer surgery}
\end{quote}

% PP2: I want them to know why. You know, her neck pain is probably from the arthritis from her surgery, cancer surgery. 

% \pquote{6}{So some of those things, you're never going to see like in orthopedic surgery, I would be so unusual to see something like coughing up blood, but that's something that we think about a lot in lung surgery...  so sort of being able to stratify some of those nuances becomes ... exciting to challenge AI and the system ... to get to that level of intelligence ...}

However, as such systems collect and leverage personal health information, some participants expressed their concerns about the information security of the system. \CP{2} mentioned that she wants to make sure she is eventually talking to people, \textit{``not some other hackers''}. Providers also discussed that some patients might be concerned about their personal information being put to malicious use, like \pquote{6}{the potential for tracking personal information}. The conflict between the expectation of personalized AI and the reservation towards AI accessing patients' personal information raises the question of balancing privacy and personalization.

% \pquote{2}{I want to make sure that it's people that I'm talking to. Not some other hackers, somebody's getting in there. }

% demographics? would that be suitable? e.g. education level, economic states
% \subsubsection{Conflicts between Automation and Human Authority}
% .

\subsubsection{Hesitation in Adoption: Trust and Power Dynamics}
\label{results:expectation}

As discussed in Section \ref{4-findings-what-and-when}, both patients and providers are excited about an AI to handle patient questions and documentation-related work automatically. However, we realize the stakeholders also have various reservations about deploying AI in the asynchronous patient-provider communication practice.

Interestingly, both groups mentioned that the other group might be unwilling to adapt to the new technology. Some providers are concerned that patients less familiar with technology and in rural communities would be too sensitive to the use of new technologies, while patients questioned whether providers would be open to \textit{``give up any of their authority to a computer''} because \epquote{6}{they've got big egos}. Some participants also disclosed their suspicion of AI-powered CAs about performance and reliability. \CP{8} revealed that she would not be interested in such technology based on her experiences, arguing that chatbots could not understand her needs correctly, while providers would ask key questions tailored to her conditions.
% P6: And so sort of overcoming some of those challenges, I think there's some people who are just always going to be unwilling to use technology, in particular, and some of our more rural communities that might be a little bit more sort of averse to medical cares in some situations and, and, and are sort of worried about the potential for tracking personal information loss, things like that. So there's just sort of a whole host of issues that that sort of come with technology with AI with everything that we do.
% \CP{6}: , but you know, what doctors? They are, they've got big egos. And they don't want to give up any of their authority to a computer.
Providers are also concerned about the system's reliability and thus suggest having healthcare providers check complicated questions and conduct counter-checks. For instance, \Pro{2} discussed the concern about misinterpretation and suggested having counter-checks.
% \begin{quote}
%     \pquote{2}{Even a system with AI is not a flawless system. They can [make] mistakes too, but sometimes we have to make sure we counter-check the system to not miss anything.} 
% \end{quote}
% \pquote{2}{.. if the patient is done Friday and saying okay, I feel some I feel bloated after eating food then we have to make sure like maybe the food is not going down right, getting stuck somewhere, taking some time to go down or I'm feeling more pain after eating food ... So sometimes the AI system may not be able to interpret those things quickly. } 

While patients and providers appreciate the automation in AI to answer their questions or take over some of their work, humans are suspicious of AI systems in actual deployment, raising questions about how humans and AI could collaborate in such communication tasks if human users do not wish to give up authority.

% wrong interpretations: P2: Any AI system can ask any patient but when those questions have to be interpreted properly, like if the patient is done Friday and saying okay, I feel some I feel bloated after eating food then we have to make sure like maybe the food is not going down right getting stuck somewhere taking some time to go down or I'm feeling more pain after eating food that need to be interpreted properly or there is some more fluid coming out of my drains. What's the color of the field that needs to be checked properly. Okay, and then a patient can say I haven't got fever with Reiger and chilled and maybe make sure like patient may be having some kind of infection that need to be taken care of immediately not to not to fit on that. So sometimes the AI system may not be able to interpret those things quickly. 

\section{Discussion}

\subsection{Design Implications on AI Systems for Patient-Provider Communication}
% why is AI better than basic health information-seeking / dumb chatbots? 
% interactivity: probe for more information, providing explanation
% flexibility / customization: protocol set by providers, different for different patients

\begin{figure}
    \centering
    \includegraphics[width=\linewidth]{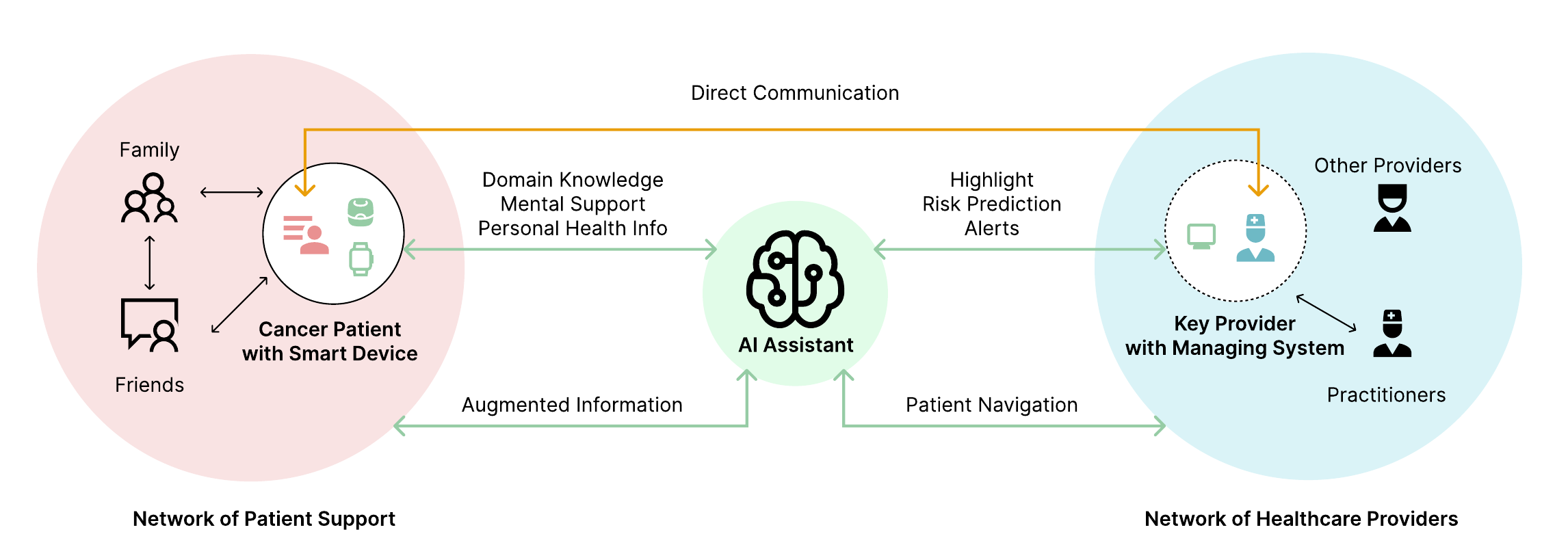}
    \caption{Overview of the AI-Enhanced Communication Paradigm for Post-Treatment Cancer Care. 1. For the cancer patient, AI could interact through user interfaces such as CAs or wearables to provide explanation and mental support using domain knowledge and collect personal health information. 2. For the key cancer care provider, AI could assist with processing information to communicate patient symptoms 3. AI manages collaborative post-treatment cancer care by augmenting medical information and navigating the patient through the healthcare network 4. Direct patient-provider communication remains crucial, empowered by AI}
    \label{fig:support-network}
\end{figure}

In this section, we provide the following design implications for future AI-powered systems to fill in the communication gaps for cancer patients and their providers based on the communication needs we summarized in Section \ref{results:overview}. We conclude our design implications to the AI-Enhanced Communication Paradigm for Post-Treatment Cancer Care in Figure \ref{fig:support-network}.

% Fig. 1. Overview of Talk2Care System. The system consists of two modules. 1) The patient module: An LLM-powered VA interface (in
% purple) that generates natural conversation with home-based older adults to collect health information and forward it to healthcare
% providers; 2) The provider module: A dashboard interface (in green) that summarizes the key information from the older patient
% conversation to assist providers who are responsible for communication (e.g., nurses and patient navigators). Note that Talk2Care
% does not provide specific healthcare advice. Our current implementation does not involve an actual electronic health record (EHR)
% system, which can be a promising future direction.

\subsubsection{Interactive AI Agents for Cancer Patients}

Based on our findings about patients' needs for explanation, reflection, and mental support, we propose that intelligent conversational agents could be a suitable interface for cancer patients. We suggest that system designers leverage the interactivity, knowledge, and ability for customization in AI, especially LLMs, to build CA, thus supporting patients in learning, communication, and mental health throughout their cancer journey.

In the post-treatment follow-up scenarios, we identified crucial information providers care about and expectations for system workflow. Future AI systems should integrate providers' protocols to lead user-AI conversations for \textbf{regular symptom checks}, focusing on crucial information and probing for details with follow-up questions to get accurate patient health information. For example, the LLM-powered system ``Talk2Care'' for older adults to communicate with their healthcare providers shows a promising direction in home-based healthcare ~\cite{yang2023talk2care}.In this process, LLMs could be particularly helpful in analyzing patient intentions and identifying risky symptoms for further action and response. The system could also provide patients with responsiveness by making the information delivery process to providers clear and transparent.

In scenarios where cancer patients have confusion or questions about their symptoms or provider instructions, LLMs could interactively \textbf{provide explanations} to answer common questions with provider documentation to support patient education. Future developers could leverage the advances in medical LLMs to expand the domain knowledge for corresponding patients~\cite{lee2020biobert,michalopoulos_umlsbert_2020,luo2022biogpt}. In cases where patients seek \textbf{mental support}, future systems may contain responses to post-treatment cancer patients considering the following three aspects. LLMs could provide explicit reassurance to patients about their treatment choices guided by providers, playing the role of supportive companion. Future work could leverage datasets in mental health to advance LLM, such as ``Mental-LLM'' ~\cite{xu2023leveraging}, to address cancer patients' emotional needs and challenges. such AI-powered systems could incorporate existing professional mental therapy or evaluation practices to monitor patients' mental health and provide interventions.

% 1. domain-specific LLM such as Mental-LLM (future)
% 2. 
With the plethora of LLMs integrated with domain knowledge, we see great potential for future AI to \textbf{support patient education}. For example, some existing work leverages LLMs to check knowledge databases for domain knowledge or verify the accuracy of the information generated by LLM and make corrections afterward \cite{yang_integrating_2023}. Although these advances in AI have been applied for professional work, there is a lack of deployment in contexts facing end users such as patients. Thus, future work may explore new contexts such as deploying AI with domain knowledge to CAs for cancer patients. In addition to these existing methods to ensure reliability in LLM, future applications of LLM in patient-provider communication contexts should also consider provider input on evaluating the reliability and having human-in-the-loop mechanisms for counter-check~\cite{cai2019hello}.

% Mental support: 

% 1. highly related to cognitive empathy, information input

% 2. expressing acknowledgment, reassurance

% 3. incorporating professional mental therapy contents/evaluation practices if possible

% \subsubsection{Management Tools for Cancer Providers}

% Our analysis reveals that cancer providers' challenges are mainly due to their busy workflow and focus on crucial information. Thus, we recommend future work leverage the advances in AI in analyzing and annotating data to design management tools for these healthcare providers, where LLMs and risk prediction models may be effective solutions. In scenarios where providers follow up with patients about updates and check-ins, systems could integrate provider protocols and inputs to capture 

% Future systems should also assist providers in effectively reviewing the human-AI communication contents, such as transcriptions, highlights, risk predictions, and alerts. In particular, while detailed information about conversations could be presented in a user interface such as a dashboard, the alerts might be integrated into different parts of the provider workflow, such as emails or mobile notifications.

\subsubsection{Connecting the Support Network}

% network moderator, server; efficiently find the right person, reduce overload
% for example, dashboard ... (can cite talk2care)

Regarding our findings on collaboration challenges and resource limitations (Section \ref{results:collaboration} and \ref{results:resource}), we advise future designers and developers of AI-powered systems to reconsider the role of AI as uniform agents that could be applicable to multiple stakeholders instead of merely being a user interface. For example, the AI agents could serve as shared assistants to support communication among the circles of support groups.

% direct to corresponding person (diagram?)

% synchronized device
We also propose that with cancer patients' vulnerability after treatments, leveraging AI in ubiquitous computing might help fill the timing gap through close monitoring and improving awareness. Researchers could consider designing such AI systems to leverage both rich personal health information and AI, thus providing a holistic view of patient health and Just-in-Time Adaptive Interventions~\cite{wang2020just,nahum2018just}. For example, AI-powered systems that include wearables and CAs might be able to detect abnormalities and communicate that information to providers.

% 1. collaboration: collect with multiple providers and support stakeholders, navigate the patients through the system with given configurations and AI analysis /  a network of supporters

% 2. import information about patient medical history, assisted by human annotations (human-AI collaboration to generate crucial information)

% 3. transparent workflow, responsiveness

% \subsection{Future AI Advances to Empower Cancer Patients and Providers}

% Through our analysis, we realize that the rarity and peril of cancer might be the essential cause of the communication challenges between cancer patients and their providers. 
% % Due to the rarity of cancer, patients lack the knowledge; combined with peril, patients are 

\subsection{Limitations and Risks of AI in Supporting Patient-Provider Communication}

Despite the great potential of AI in supporting healthcare communication, we maintain that human-human communication remains essential in post-treatment cancer care scenarios, considering the severity, complexity, and high risk of cancer and post-treatment complications. Providers' engagement remains irreplaceable since the professionals had to decide on essential treatment plans and complicated situations. In key clinical diagnosis and decision practices, direct patient-provider communication, preferably synchronous communication, should still be emphasized to ensure accurate understanding and informed decisions. Considering cancer patients' mental needs and reservations towards AI, providers' human response throughout such AI-powered systems also remains essential in their patient-provider communication.

In addition, AI is not necessarily the best solution for simple communication tasks such as prescription requests or scheduling appointments, considering the cost and risks in development and deployment. Firstly, such AI-powered systems inevitably have privacy and security risks as they store and utilize personal health information, medical history, and possibly other identifiable information. Apart from ensuring the patient data are stored in secure databases that follow HIPPA compliance~\cite{rightsocrHealthInformationPrivacy2021}, system developers should also implement strategies to make use of minimal personal information, such as using Natural Language Processing methods, to filter keywords and providing reminders or warnings to users ~\cite{gangavarapu2020applicability, subramani2023detecting}. 

Meanwhile, regarding reliability risks, whenever the responses of AI systems involve interpreting medical knowledge, clinical instructions, or performing risk predictions, such systems should be under close supervision of professionals such as healthcare providers to ensure the system does not provide false information. A human-AI collaboration workflow should be established, such as clinical communication practices in post-treatment cancer care.

% some reliability risks; future work may need experts and users to evaluate risk

\section{Limitations}

% 1. not specific cancer, they have different complications and treatments just general cancer
% 2. number, hard to recruit (similar to 1)
% 3. patient-provider pair
% 4. LLM -> actually ... ; did not reon their trustworthiness and ethical concern ; beyond the scope

Our study has a relatively limited number of participants since cancer providers and survivors who had post-treatment communication with their providers are relatively difficult to recruit. For providers, we mainly recruited surgeons to focus on health concerns, and have comparatively limited insights from other healthcare providers and support staff. Thus, based on our findings, future work could further explore the communication challenges within providers' collaboration. We also recognize we did not recruit providers and patients of a specific cancer type or treatment type, or recruit patient-provider pairs. Since participants had different cancer treatments and post-treatment complications, our findings are more generalized to cancer populations. Future work may explore in-depth the communication challenges and AI opportunities of a specific type of cancer and specific treatments with a greater number of patient-provider pairs, which is beyond the scope of our study.

% Also, we found no significant use of novel applications for cancer patients. Yet, with prior work designing mobile applications for cancer patients or health monitoring systems, the communication challenges in those specific user communities could differ from our study. % may have specific needs, we will consider in our future work

Regarding our RQs about AI potentials, we discussed AI trustworthiness and ethical concerns but did not probe in-depth into participants' attitudes toward such AI risks, since we primarily aim to explore AI potentials by addressing communication challenges. We also recognize that we had open discussions with our participants about future AI technologies but did not have them interact with such a system directly during this study. Thus, the following work of this study may consider designing and deploying such an AI-powered system to evaluate the design implications. Given the high risk in post-treatment cancer care, future work could further discuss with stakeholders the trustworthiness and ethical concerns of AI in their patient-provider communication.

% generalizability: may generalize to other patients with serious illness / need long-term monitoring; may not generalize to other illnesses
\section{Conclusion}
Through our inquiry into patient-provider communication practices, the present paper identifies and analyzes the challenges in remote communication between cancer patients and their healthcare providers. Moreover, based on user expectations for AI technologies for communication throughout the cancer journey, this work provides design implications for future work supporting patient-provider communication with AI. 
% knowledge
% timing
% emotional
% collaboration
% resource and accessibility
% personalization and customization
Cancer patients and providers experience communication challenges due to their gaps in knowledge, timing needs, and emotional focus. With the rarity, severity, and variety of cancer, these gaps also result in challenges in collaboration and navigating resources in patient-provider communication. Thus, we propose that future system designers and researchers could leverage AI to support their communication by designing LLM-powered CA to provide patients with related knowledge and expected responsiveness. We also suggest utilizing AI as a provider management tool while picturing multi-agent systems for collaboration in the support network. 
We encourage future AI advances to work on providing reliable domain knowledge under professional instructions and explore novel multi-modality AI systems to further address the communication challenges.

% "Our research highlights the significant potential of AI in transforming post-treatment cancer care communication. We identified key gaps in patient-provider interactions and proposed AI as a tool to bridge these gaps, offering customized, timely, and efficient communication. However, this comes with challenges in maintaining the balance between personalization and privacy, as well as ensuring the seamless integration of AI into existing healthcare workflows. The future of healthcare communication lies in a collaborative model where AI complements human expertise, ensuring a patient-centered approach while respecting ethical boundaries. Our study paves the way for further exploration into AI-enhanced healthcare solutions, aiming for a future where technology and human compassion work hand in hand for improved patient outcomes."

%%
%% The acknowledgments section is defined using the "acks" environment
%% (and NOT an unnumbered section). This ensures the proper
%% identification of the section in the article metadata, and the
%% consistent spelling of the heading.
% \begin{acks}
% To Robert, for the bagels and explaining CMYK and color spaces.
% \end{acks}

%%
%% The next two lines define the bibliography style to be used, and
%% the bibliography file.
\bibliographystyle{ACM-Reference-Format}
\bibliography{
bib/1-multi-communication,
bib/2-agents-LLM,
bib/3-commu-LLM,
bib/11-cancer-care,
bib/12-communication-technology,
bib/13-AI,
bib/20-other
}

%%% -*-BibTeX-*-
%%% Do NOT edit. File created by BibTeX with style
%%% ACM-Reference-Format-Journals [18-Jan-2012].

\begin{thebibliography}{79}

%%% ====================================================================
%%% NOTE TO THE USER: you can override these defaults by providing
%%% customized versions of any of these macros before the \bibliography
%%% command.  Each of them MUST provide its own final punctuation,
%%% except for \shownote{}, \showDOI{}, and \showURL{}.  The latter two
%%% do not use final punctuation, in order to avoid confusing it with
%%% the Web address.
%%%
%%% To suppress output of a particular field, define its macro to expand
%%% to an empty string, or better, \unskip, like this:
%%%
%%% \newcommand{\showDOI}[1]{\unskip}   % LaTeX syntax
%%%
%%% \def \showDOI #1{\unskip}           % plain TeX syntax
%%%
%%% ====================================================================

\ifx \showCODEN    \undefined \def \showCODEN     #1{\unskip}     \fi
\ifx \showDOI      \undefined \def \showDOI       #1{#1}\fi
\ifx \showISBNx    \undefined \def \showISBNx     #1{\unskip}     \fi
\ifx \showISBNxiii \undefined \def \showISBNxiii  #1{\unskip}     \fi
\ifx \showISSN     \undefined \def \showISSN      #1{\unskip}     \fi
\ifx \showLCCN     \undefined \def \showLCCN      #1{\unskip}     \fi
\ifx \shownote     \undefined \def \shownote      #1{#1}          \fi
\ifx \showarticletitle \undefined \def \showarticletitle #1{#1}   \fi
\ifx \showURL      \undefined \def \showURL       {\relax}        \fi
% The following commands are used for tagged output and should be
% invisible to TeX
\providecommand\bibfield[2]{#2}
\providecommand\bibinfo[2]{#2}
\providecommand\natexlab[1]{#1}
\providecommand\showeprint[2][]{arXiv:#2}

\bibitem[noa(2023)]%
        {noauthor_telehealth_nodate}
 \bibinfo{year}{2023}\natexlab{}.
\newblock \bibinfo{title}{Telehealth: {Technology} meets health care - {Mayo} {Clinic}}.
\newblock
\newblock
\urldef\tempurl%
\url{https://www.mayoclinic.org/healthy-lifestyle/consumer-health/in-depth/telehealth/art-20044878}
\showURL{%
\tempurl}


\bibitem[Akram et~al\mbox{.}(2017)]%
        {akram2017awareness}
\bibfield{author}{\bibinfo{person}{Muhammad Akram}, \bibinfo{person}{Mehwish Iqbal}, \bibinfo{person}{Muhammad Daniyal}, {and} \bibinfo{person}{Asmat~Ullah Khan}.} \bibinfo{year}{2017}\natexlab{}.
\newblock \showarticletitle{Awareness and current knowledge of breast cancer}.
\newblock \bibinfo{journal}{\emph{Biological research}}  \bibinfo{volume}{50} (\bibinfo{year}{2017}), \bibinfo{pages}{1--23}.
\newblock


\bibitem[Ali et~al\mbox{.}(2023)]%
        {ali2023using}
\bibfield{author}{\bibinfo{person}{Stephen~R Ali}, \bibinfo{person}{Thomas~D Dobbs}, \bibinfo{person}{Hayley~A Hutchings}, {and} \bibinfo{person}{Iain~S Whitaker}.} \bibinfo{year}{2023}\natexlab{}.
\newblock \showarticletitle{Using ChatGPT to write patient clinic letters}.
\newblock \bibinfo{journal}{\emph{The Lancet Digital Health}} \bibinfo{volume}{5}, \bibinfo{number}{4} (\bibinfo{year}{2023}), \bibinfo{pages}{e179--e181}.
\newblock


\bibitem[Andy et~al\mbox{.}(2021)]%
        {andy2021understanding}
\bibfield{author}{\bibinfo{person}{Anietie Andy}, \bibinfo{person}{Uduak Andy}, {et~al\mbox{.}}} \bibinfo{year}{2021}\natexlab{}.
\newblock \showarticletitle{Understanding communication in an online cancer forum: content analysis study}.
\newblock \bibinfo{journal}{\emph{JMIR cancer}} \bibinfo{volume}{7}, \bibinfo{number}{3} (\bibinfo{year}{2021}), \bibinfo{pages}{e29555}.
\newblock


\bibitem[Anil et~al\mbox{.}(2023)]%
        {anil2023palm}
\bibfield{author}{\bibinfo{person}{Rohan Anil}, \bibinfo{person}{Andrew~M Dai}, \bibinfo{person}{Orhan Firat}, \bibinfo{person}{Melvin Johnson}, \bibinfo{person}{Dmitry Lepikhin}, \bibinfo{person}{Alexandre Passos}, \bibinfo{person}{Siamak Shakeri}, \bibinfo{person}{Emanuel Taropa}, \bibinfo{person}{Paige Bailey}, \bibinfo{person}{Zhifeng Chen}, {et~al\mbox{.}}} \bibinfo{year}{2023}\natexlab{}.
\newblock \showarticletitle{Palm 2 technical report}.
\newblock \bibinfo{journal}{\emph{arXiv preprint arXiv:2305.10403}} (\bibinfo{year}{2023}).
\newblock


\bibitem[Beltagy et~al\mbox{.}(2019)]%
        {beltagy2019scibert}
\bibfield{author}{\bibinfo{person}{Iz Beltagy}, \bibinfo{person}{Kyle Lo}, {and} \bibinfo{person}{Arman Cohan}.} \bibinfo{year}{2019}\natexlab{}.
\newblock \showarticletitle{SciBERT: A pretrained language model for scientific text}.
\newblock \bibinfo{journal}{\emph{arXiv preprint arXiv:1903.10676}} (\bibinfo{year}{2019}).
\newblock


\bibitem[Bhat et~al\mbox{.}(2021)]%
        {bhat_infrastructuring_2021}
\bibfield{author}{\bibinfo{person}{Karthik~S Bhat}, \bibinfo{person}{Mohit Jain}, {and} \bibinfo{person}{Neha Kumar}.} \bibinfo{year}{2021}\natexlab{}.
\newblock \showarticletitle{Infrastructuring {Telehealth} in ({In}){Formal} {Patient}-{Doctor} {Contexts}}.
\newblock \bibinfo{journal}{\emph{Proceedings of the ACM on Human-Computer Interaction}} \bibinfo{volume}{5}, \bibinfo{number}{CSCW2} (\bibinfo{date}{Oct.} \bibinfo{year}{2021}), \bibinfo{pages}{1--28}.
\newblock
\showISSN{2573-0142}
\urldef\tempurl%
\url{https://doi.org/10.1145/3476064}
\showDOI{\tempurl}


\bibitem[Braun and Clarke(2006)]%
        {braun2006using}
\bibfield{author}{\bibinfo{person}{Virginia Braun} {and} \bibinfo{person}{Victoria Clarke}.} \bibinfo{year}{2006}\natexlab{}.
\newblock \showarticletitle{Using thematic analysis in psychology}.
\newblock \bibinfo{journal}{\emph{Qualitative research in psychology}} \bibinfo{volume}{3}, \bibinfo{number}{2} (\bibinfo{year}{2006}), \bibinfo{pages}{77}.
\newblock


\bibitem[Cai et~al\mbox{.}(2019)]%
        {cai2019hello}
\bibfield{author}{\bibinfo{person}{Carrie~J Cai}, \bibinfo{person}{Samantha Winter}, \bibinfo{person}{David Steiner}, \bibinfo{person}{Lauren Wilcox}, {and} \bibinfo{person}{Michael Terry}.} \bibinfo{year}{2019}\natexlab{}.
\newblock \showarticletitle{" Hello AI": uncovering the onboarding needs of medical practitioners for human-AI collaborative decision-making}.
\newblock \bibinfo{journal}{\emph{Proceedings of the ACM on Human-computer Interaction}} \bibinfo{volume}{3}, \bibinfo{number}{CSCW} (\bibinfo{year}{2019}), \bibinfo{pages}{1--24}.
\newblock


\bibitem[Chan et~al\mbox{.}(2023)]%
        {chan_chateval_2023}
\bibfield{author}{\bibinfo{person}{Chi-Min Chan}, \bibinfo{person}{Weize Chen}, \bibinfo{person}{Yusheng Su}, \bibinfo{person}{Jianxuan Yu}, \bibinfo{person}{Wei Xue}, \bibinfo{person}{Shanghang Zhang}, \bibinfo{person}{Jie Fu}, {and} \bibinfo{person}{Zhiyuan Liu}.} \bibinfo{year}{2023}\natexlab{}.
\newblock \bibinfo{title}{{ChatEval}: {Towards} {Better} {LLM}-based {Evaluators} through {Multi}-{Agent} {Debate}}.
\newblock
\newblock
\urldef\tempurl%
\url{http://arxiv.org/abs/2308.07201}
\showURL{%
\tempurl}
\newblock
\shownote{arXiv:2308.07201 [cs]}.


\bibitem[Chandwani and Kumar(2018)]%
        {chandwani_stitching_2018}
\bibfield{author}{\bibinfo{person}{Rajesh Chandwani} {and} \bibinfo{person}{Neha Kumar}.} \bibinfo{year}{2018}\natexlab{}.
\newblock \showarticletitle{Stitching {Infrastructures} to {Facilitate} {Telemedicine} for {Low}-{Resource} {Environments}}. In \bibinfo{booktitle}{\emph{Proceedings of the 2018 {CHI} {Conference} on {Human} {Factors} in {Computing} {Systems}}}. \bibinfo{publisher}{ACM}, \bibinfo{address}{Montreal QC Canada}, \bibinfo{pages}{1--12}.
\newblock
\showISBNx{978-1-4503-5620-6}
\urldef\tempurl%
\url{https://doi.org/10.1145/3173574.3173958}
\showDOI{\tempurl}


\bibitem[Ch{\'a}varri-Guerra et~al\mbox{.}(2021)]%
        {chavarri2021providing}
\bibfield{author}{\bibinfo{person}{Yanin Ch{\'a}varri-Guerra}, \bibinfo{person}{Wendy~Alicia Ramos-L{\'o}pez}, \bibinfo{person}{Alfredo Covarrubias-G{\'o}mez}, \bibinfo{person}{Sof{\'\i}a S{\'a}nchez-Rom{\'a}n}, \bibinfo{person}{Paulina Quiroz-Friedman}, \bibinfo{person}{Natasha Alcocer-Castillejos}, \bibinfo{person}{Mar{\'\i}a del Pilar Milke-Garc{\'\i}a}, \bibinfo{person}{M{\'o}nica Carrillo-Soto}, \bibinfo{person}{Andrea Morales-Alfaro}, \bibinfo{person}{Mildred Medina-Palma}, {et~al\mbox{.}}} \bibinfo{year}{2021}\natexlab{}.
\newblock \showarticletitle{Providing supportive and palliative care using telemedicine for patients with advanced cancer during the COVID-19 pandemic in Mexico}.
\newblock \bibinfo{journal}{\emph{The oncologist}} \bibinfo{volume}{26}, \bibinfo{number}{3} (\bibinfo{year}{2021}), \bibinfo{pages}{e512--e515}.
\newblock


\bibitem[Chen et~al\mbox{.}(2023)]%
        {chen_llm-empowered_2023}
\bibfield{author}{\bibinfo{person}{Siyuan Chen}, \bibinfo{person}{Mengyue Wu}, \bibinfo{person}{Kenny~Q. Zhu}, \bibinfo{person}{Kunyao Lan}, \bibinfo{person}{Zhiling Zhang}, {and} \bibinfo{person}{Lyuchun Cui}.} \bibinfo{year}{2023}\natexlab{}.
\newblock \bibinfo{title}{{LLM}-empowered {Chatbots} for {Psychiatrist} and {Patient} {Simulation}: {Application} and {Evaluation}}.
\newblock
\newblock
\urldef\tempurl%
\url{http://arxiv.org/abs/2305.13614}
\showURL{%
\tempurl}
\newblock
\shownote{arXiv:2305.13614 [cs]}.


\bibitem[Chung et~al\mbox{.}(2015)]%
        {chung_more_2015}
\bibfield{author}{\bibinfo{person}{Chia-Fang Chung}, \bibinfo{person}{Jonathan Cook}, \bibinfo{person}{Elizabeth Bales}, \bibinfo{person}{Jasmine Zia}, {and} \bibinfo{person}{Sean~A Munson}.} \bibinfo{year}{2015}\natexlab{}.
\newblock \showarticletitle{More {Than} {Telemonitoring}: {Health} {Provider} {Use} and {Nonuse} of {Life}-{Log} {Data} in {Irritable} {Bowel} {Syndrome} and {Weight} {Management}}.
\newblock \bibinfo{journal}{\emph{Journal of Medical Internet Research}} \bibinfo{volume}{17}, \bibinfo{number}{8} (\bibinfo{date}{Aug.} \bibinfo{year}{2015}), \bibinfo{pages}{e203}.
\newblock
\showISSN{1438-8871}
\urldef\tempurl%
\url{https://doi.org/10.2196/jmir.4364}
\showDOI{\tempurl}


\bibitem[Clark and Kelliher(2021)]%
        {clark_understanding_2021}
\bibfield{author}{\bibinfo{person}{Juliet Clark} {and} \bibinfo{person}{Aisling Kelliher}.} \bibinfo{year}{2021}\natexlab{}.
\newblock \showarticletitle{Understanding the {Needs} and {Values} of {Rehabilitation} {Therapists} in {Designing} and {Implementing} {Telehealth} {Solutions}}. In \bibinfo{booktitle}{\emph{Extended {Abstracts} of the 2021 {CHI} {Conference} on {Human} {Factors} in {Computing} {Systems}}}. \bibinfo{publisher}{ACM}, \bibinfo{address}{Yokohama Japan}, \bibinfo{pages}{1--6}.
\newblock
\showISBNx{978-1-4503-8095-9}
\urldef\tempurl%
\url{https://doi.org/10.1145/3411763.3451704}
\showDOI{\tempurl}


\bibitem[Corbin and Strauss(1990)]%
        {corbin1990grounded}
\bibfield{author}{\bibinfo{person}{Juliet~M Corbin} {and} \bibinfo{person}{Anselm Strauss}.} \bibinfo{year}{1990}\natexlab{}.
\newblock \showarticletitle{Grounded theory research: Procedures, canons, and evaluative criteria}.
\newblock \bibinfo{journal}{\emph{Qualitative sociology}} \bibinfo{volume}{13}, \bibinfo{number}{1} (\bibinfo{year}{1990}), \bibinfo{pages}{3--21}.
\newblock


\bibitem[Donelan et~al\mbox{.}(2019)]%
        {donelan_patient_2019}
\bibfield{author}{\bibinfo{person}{Karen Donelan}, \bibinfo{person}{Esteban~A Barreto}, \bibinfo{person}{Sarah Sossong}, \bibinfo{person}{Carie Michael}, \bibinfo{person}{Juan~J Estrada}, \bibinfo{person}{Adam~B Cohen}, \bibinfo{person}{Janet Wozniak}, {and} \bibinfo{person}{Lee~H Schwamm}.} \bibinfo{year}{2019}\natexlab{}.
\newblock \showarticletitle{Patient and {Clinician} {Experiences} {With} {Telehealth} for {Patient} {Follow}-up {Care}}.
\newblock \bibinfo{journal}{\emph{THE AMERICAN JOURNAL OF MANAGED CARE}} (\bibinfo{year}{2019}).
\newblock


\bibitem[Fang et~al\mbox{.}(2023)]%
        {fang_socializechat_2023}
\bibfield{author}{\bibinfo{person}{Yuyang Fang}, \bibinfo{person}{Yunkai Xu}, \bibinfo{person}{Zhuyu Teng}, \bibinfo{person}{Zhaoqu Jiang}, {and} \bibinfo{person}{Wei Xiang}.} \bibinfo{year}{2023}\natexlab{}.
\newblock \showarticletitle{{SocializeChat}: a {GPT}-based {AAC} {Tool} for {Social} {Communication} {Through} {Eye} {Gazing}}. In \bibinfo{booktitle}{\emph{Adjunct {Proceedings} of the 2023 {ACM} {International} {Joint} {Conference} on {Pervasive} and {Ubiquitous} {Computing} \& the 2023 {ACM} {International} {Symposium} on {Wearable} {Computing}}}. \bibinfo{publisher}{ACM}, \bibinfo{address}{Cancun, Quintana Roo Mexico}, \bibinfo{pages}{128--132}.
\newblock
\showISBNx{9798400702006}
\urldef\tempurl%
\url{https://doi.org/10.1145/3594739.3610705}
\showDOI{\tempurl}


\bibitem[Ferlay et~al\mbox{.}(2021)]%
        {ferlay2021cancer}
\bibfield{author}{\bibinfo{person}{Jacques Ferlay}, \bibinfo{person}{Murielle Colombet}, \bibinfo{person}{Isabelle Soerjomataram}, \bibinfo{person}{Donald~M Parkin}, \bibinfo{person}{Marion Pi{\~n}eros}, \bibinfo{person}{Ariana Znaor}, {and} \bibinfo{person}{Freddie Bray}.} \bibinfo{year}{2021}\natexlab{}.
\newblock \showarticletitle{Cancer statistics for the year 2020: An overview}.
\newblock \bibinfo{journal}{\emph{International journal of cancer}} \bibinfo{volume}{149}, \bibinfo{number}{4} (\bibinfo{year}{2021}), \bibinfo{pages}{778--789}.
\newblock


\bibitem[Frenkel et~al\mbox{.}(2016)]%
        {frenkel2016exceptional}
\bibfield{author}{\bibinfo{person}{Moshe Frenkel}, \bibinfo{person}{Joan~C Engebretson}, \bibinfo{person}{Sky Gross}, \bibinfo{person}{Noemi~E Peterson}, \bibinfo{person}{Ariela~Popper Giveon}, \bibinfo{person}{Kenneth Sapire}, {and} \bibinfo{person}{Doron Hermoni}.} \bibinfo{year}{2016}\natexlab{}.
\newblock \showarticletitle{Exceptional patients and communication in cancer care—are we missing another survival factor?}
\newblock \bibinfo{journal}{\emph{Supportive Care in Cancer}}  \bibinfo{volume}{24} (\bibinfo{year}{2016}), \bibinfo{pages}{4249--4255}.
\newblock


\bibitem[Gangavarapu et~al\mbox{.}(2020)]%
        {gangavarapu2020applicability}
\bibfield{author}{\bibinfo{person}{Tushaar Gangavarapu}, \bibinfo{person}{CD Jaidhar}, {and} \bibinfo{person}{Bhabesh Chanduka}.} \bibinfo{year}{2020}\natexlab{}.
\newblock \showarticletitle{Applicability of machine learning in spam and phishing email filtering: review and approaches}.
\newblock \bibinfo{journal}{\emph{Artificial Intelligence Review}}  \bibinfo{volume}{53} (\bibinfo{year}{2020}), \bibinfo{pages}{5019--5081}.
\newblock


\bibitem[Gilson et~al\mbox{.}(2023)]%
        {gilson2023does}
\bibfield{author}{\bibinfo{person}{Aidan Gilson}, \bibinfo{person}{Conrad~W Safranek}, \bibinfo{person}{Thomas Huang}, \bibinfo{person}{Vimig Socrates}, \bibinfo{person}{Ling Chi}, \bibinfo{person}{Richard~Andrew Taylor}, \bibinfo{person}{David Chartash}, {et~al\mbox{.}}} \bibinfo{year}{2023}\natexlab{}.
\newblock \showarticletitle{How does ChatGPT perform on the United States medical licensing examination? The implications of large language models for medical education and knowledge assessment}.
\newblock \bibinfo{journal}{\emph{JMIR Medical Education}} \bibinfo{volume}{9}, \bibinfo{number}{1} (\bibinfo{year}{2023}), \bibinfo{pages}{e45312}.
\newblock


\bibitem[Glaser et~al\mbox{.}(1968)]%
        {glaser1968discovery}
\bibfield{author}{\bibinfo{person}{Barney~G Glaser}, \bibinfo{person}{Anselm~L Strauss}, {and} \bibinfo{person}{Elizabeth Strutzel}.} \bibinfo{year}{1968}\natexlab{}.
\newblock \showarticletitle{The discovery of grounded theory; strategies for qualitative research}.
\newblock \bibinfo{journal}{\emph{Nursing research}} \bibinfo{volume}{17}, \bibinfo{number}{4} (\bibinfo{year}{1968}), \bibinfo{pages}{364}.
\newblock


\bibitem[Gon{\c c}alves-Bradley et~al\mbox{.}(2020)]%
        {goncalves-bradley_mobile_2020}
\bibfield{author}{\bibinfo{person}{Daniela~C. Gon{\c c}alves-Bradley}, \bibinfo{person}{Ana Rita~J. Maria}, \bibinfo{person}{Ignacio Ricci-Cabello}, \bibinfo{person}{Gemma Villanueva}, \bibinfo{person}{Marita~S. F{\o}nhus}, \bibinfo{person}{Claire Glenton}, \bibinfo{person}{Simon Lewin}, \bibinfo{person}{Nicholas Henschke}, \bibinfo{person}{Brian~S. Buckley}, \bibinfo{person}{Garrett~L. Mehl}, \bibinfo{person}{Tigest Tamrat}, {and} \bibinfo{person}{Sasha Shepperd}.} \bibinfo{year}{2020}\natexlab{}.
\newblock \showarticletitle{Mobile technologies to support healthcare provider to healthcare provider communication and management of care}.
\newblock \bibinfo{journal}{\emph{Cochrane Database of Systematic Reviews}} \bibinfo{number}{8} (\bibinfo{year}{2020}).
\newblock
\showISSN{1465-1858}
\urldef\tempurl%
\url{https://doi.org/10.1002/14651858.CD012927.pub2}
\showDOI{\tempurl}
\newblock
\shownote{Publisher: John Wiley \& Sons, Ltd}.


\bibitem[Gupta et~al\mbox{.}(2015)]%
        {gupta2015review}
\bibfield{author}{\bibinfo{person}{Addya Gupta}, \bibinfo{person}{K Shridhar}, {and} \bibinfo{person}{PK Dhillon}.} \bibinfo{year}{2015}\natexlab{}.
\newblock \showarticletitle{A review of breast cancer awareness among women in India: Cancer literate or awareness deficit?}
\newblock \bibinfo{journal}{\emph{European Journal of Cancer}} \bibinfo{volume}{51}, \bibinfo{number}{14} (\bibinfo{year}{2015}), \bibinfo{pages}{2058--2066}.
\newblock


\bibitem[Heng et~al\mbox{.}(2011)]%
        {heng2011evolutionary}
\bibfield{author}{\bibinfo{person}{Henry~HQ Heng}, \bibinfo{person}{Joshua~B Stevens}, \bibinfo{person}{Steven~W Bremer}, \bibinfo{person}{Guo Liu}, \bibinfo{person}{Batoul~Y Abdallah}, {and} \bibinfo{person}{J~Ye Christine}.} \bibinfo{year}{2011}\natexlab{}.
\newblock \showarticletitle{Evolutionary mechanisms and diversity in cancer}.
\newblock \bibinfo{journal}{\emph{Advances in cancer research}}  \bibinfo{volume}{112} (\bibinfo{year}{2011}), \bibinfo{pages}{217--253}.
\newblock


\bibitem[Hong et~al\mbox{.}(2020)]%
        {hong_digital_2020}
\bibfield{author}{\bibinfo{person}{Y.~Alicia Hong}, \bibinfo{person}{Md~Mahbub Hossain}, {and} \bibinfo{person}{Wen-Ying~Sylvia Chou}.} \bibinfo{year}{2020}\natexlab{}.
\newblock \showarticletitle{Digital interventions to facilitate patient-provider communication in cancer care: {A} systematic review}.
\newblock \bibinfo{journal}{\emph{Psycho-Oncology}} \bibinfo{volume}{29}, \bibinfo{number}{4} (\bibinfo{year}{2020}), \bibinfo{pages}{591--603}.
\newblock
\showISSN{1099-1611}
\urldef\tempurl%
\url{https://doi.org/10.1002/pon.5310}
\showDOI{\tempurl}
\newblock
\shownote{\_eprint: https://onlinelibrary.wiley.com/doi/pdf/10.1002/pon.5310}.


\bibitem[Ilic and Ilic(2016)]%
        {ilic2016epidemiology}
\bibfield{author}{\bibinfo{person}{Milena Ilic} {and} \bibinfo{person}{Irena Ilic}.} \bibinfo{year}{2016}\natexlab{}.
\newblock \showarticletitle{Epidemiology of pancreatic cancer}.
\newblock \bibinfo{journal}{\emph{World journal of gastroenterology}} \bibinfo{volume}{22}, \bibinfo{number}{44} (\bibinfo{year}{2016}), \bibinfo{pages}{9694}.
\newblock


\bibitem[Iqbal et~al\mbox{.}(2015)]%
        {iqbal2015differences}
\bibfield{author}{\bibinfo{person}{Javaid Iqbal}, \bibinfo{person}{Ophira Ginsburg}, \bibinfo{person}{Paula~A Rochon}, \bibinfo{person}{Ping Sun}, {and} \bibinfo{person}{Steven~A Narod}.} \bibinfo{year}{2015}\natexlab{}.
\newblock \showarticletitle{Differences in breast cancer stage at diagnosis and cancer-specific survival by race and ethnicity in the United States}.
\newblock \bibinfo{journal}{\emph{Jama}} \bibinfo{volume}{313}, \bibinfo{number}{2} (\bibinfo{year}{2015}), \bibinfo{pages}{165--173}.
\newblock


\bibitem[Jacobs et~al\mbox{.}(2014)]%
        {jacobs2014cancer}
\bibfield{author}{\bibinfo{person}{Maia Jacobs}, \bibinfo{person}{James Clawson}, {and} \bibinfo{person}{Elizabeth~D Mynatt}.} \bibinfo{year}{2014}\natexlab{}.
\newblock \showarticletitle{Cancer navigation: opportunities and challenges for facilitating the breast cancer journey}. In \bibinfo{booktitle}{\emph{Proceedings of the 17th ACM conference on Computer supported cooperative work \& social computing}}. \bibinfo{pages}{1467--1478}.
\newblock


\bibitem[Jacobs et~al\mbox{.}(2018)]%
        {jacobsMyPathInvestigatingBreast2018}
\bibfield{author}{\bibinfo{person}{Maia Jacobs}, \bibinfo{person}{Jeremy Johnson}, {and} \bibinfo{person}{Elizabeth~D. Mynatt}.} \bibinfo{year}{2018}\natexlab{}.
\newblock \showarticletitle{{{MyPath}}: {{Investigating Breast Cancer Patients}}' {{Use}} of {{Personalized Health Information}}}.
\newblock \bibinfo{journal}{\emph{Proceedings of the ACM on Human-Computer Interaction}} \bibinfo{volume}{2}, \bibinfo{number}{CSCW} (\bibinfo{date}{Nov.} \bibinfo{year}{2018}), \bibinfo{pages}{78:1--78:21}.
\newblock
\urldef\tempurl%
\url{https://doi.org/10.1145/3274347}
\showDOI{\tempurl}


\bibitem[Jacobs et~al\mbox{.}(2015)]%
        {jacobs_comparing_2015}
\bibfield{author}{\bibinfo{person}{Maia~L. Jacobs}, \bibinfo{person}{James Clawson}, {and} \bibinfo{person}{Elizabeth~D. Mynatt}.} \bibinfo{year}{2015}\natexlab{}.
\newblock \showarticletitle{Comparing {Health} {Information} {Sharing} {Preferences} of {Cancer} {Patients}, {Doctors}, and {Navigators}}. In \bibinfo{booktitle}{\emph{Proceedings of the 18th {ACM} {Conference} on {Computer} {Supported} {Cooperative} {Work} \& {Social} {Computing}}} \emph{(\bibinfo{series}{{CSCW} '15})}. \bibinfo{publisher}{Association for Computing Machinery}, \bibinfo{address}{New York, NY, USA}, \bibinfo{pages}{808--818}.
\newblock
\showISBNx{978-1-4503-2922-4}
\urldef\tempurl%
\url{https://doi.org/10.1145/2675133.2675252}
\showDOI{\tempurl}


\bibitem[Johnson et~al\mbox{.}(2023)]%
        {johnson2023assessing}
\bibfield{author}{\bibinfo{person}{Douglas Johnson}, \bibinfo{person}{Rachel Goodman}, \bibinfo{person}{J Patrinely}, \bibinfo{person}{Cosby Stone}, \bibinfo{person}{Eli Zimmerman}, \bibinfo{person}{Rebecca Donald}, \bibinfo{person}{Sam Chang}, \bibinfo{person}{Sean Berkowitz}, \bibinfo{person}{Avni Finn}, \bibinfo{person}{Eiman Jahangir}, {et~al\mbox{.}}} \bibinfo{year}{2023}\natexlab{}.
\newblock \showarticletitle{Assessing the accuracy and reliability of AI-generated medical responses: an evaluation of the Chat-GPT model}.
\newblock \bibinfo{journal}{\emph{Research square}} (\bibinfo{year}{2023}).
\newblock


\bibitem[Jones et~al\mbox{.}(2016)]%
        {jones2016cancer}
\bibfield{author}{\bibinfo{person}{Jennifer~M Jones}, \bibinfo{person}{Karin Olson}, \bibinfo{person}{Pamela Catton}, \bibinfo{person}{Charles~N Catton}, \bibinfo{person}{Neil~E Fleshner}, \bibinfo{person}{Monika~K Krzyzanowska}, \bibinfo{person}{David~R McCready}, \bibinfo{person}{Rebecca~KS Wong}, \bibinfo{person}{Haiyan Jiang}, {and} \bibinfo{person}{Doris Howell}.} \bibinfo{year}{2016}\natexlab{}.
\newblock \showarticletitle{Cancer-related fatigue and associated disability in post-treatment cancer survivors}.
\newblock \bibinfo{journal}{\emph{Journal of Cancer Survivorship}}  \bibinfo{volume}{10} (\bibinfo{year}{2016}), \bibinfo{pages}{51--61}.
\newblock


\bibitem[Kabir et~al\mbox{.}(2019)]%
        {kabir2019m}
\bibfield{author}{\bibinfo{person}{Kazi~Sinthia Kabir}, \bibinfo{person}{Erin~L Van~Blarigan}, \bibinfo{person}{June~M Chan}, \bibinfo{person}{Stacey~A Kenfield}, {and} \bibinfo{person}{Jason Wiese}.} \bibinfo{year}{2019}\natexlab{}.
\newblock \showarticletitle{" I'm Done with Cancer. What am I Trying to Improve?" Understanding the Perspective of Prostate Cancer Patients to Support Multiple Health Behavior Change}. In \bibinfo{booktitle}{\emph{Proceedings of the 13th EAI International Conference on Pervasive Computing Technologies for Healthcare}}. \bibinfo{pages}{81--90}.
\newblock


\bibitem[Karusala et~al\mbox{.}(2020)]%
        {karusala_making_2020}
\bibfield{author}{\bibinfo{person}{Naveena Karusala}, \bibinfo{person}{Ding Wang}, {and} \bibinfo{person}{Jacki O'Neill}.} \bibinfo{year}{2020}\natexlab{}.
\newblock \showarticletitle{Making {Chat} at {Home} in the {Hospital}: {Exploring} {Chat} {Use} by {Nurses}}. In \bibinfo{booktitle}{\emph{Proceedings of the 2020 {CHI} {Conference} on {Human} {Factors} in {Computing} {Systems}}}. \bibinfo{publisher}{ACM}, \bibinfo{address}{Honolulu HI USA}, \bibinfo{pages}{1--15}.
\newblock
\showISBNx{978-1-4503-6708-0}
\urldef\tempurl%
\url{https://doi.org/10.1145/3313831.3376166}
\showDOI{\tempurl}


\bibitem[Kent et~al\mbox{.}(2016)]%
        {kent2016caring}
\bibfield{author}{\bibinfo{person}{Erin~E Kent}, \bibinfo{person}{Julia~H Rowland}, \bibinfo{person}{Laurel Northouse}, \bibinfo{person}{Kristin Litzelman}, \bibinfo{person}{Wen-Ying~Sylvia Chou}, \bibinfo{person}{Nonniekaye Shelburne}, \bibinfo{person}{Catherine Timura}, \bibinfo{person}{Ann O'Mara}, {and} \bibinfo{person}{Karen Huss}.} \bibinfo{year}{2016}\natexlab{}.
\newblock \showarticletitle{Caring for caregivers and patients: research and clinical priorities for informal cancer caregiving}.
\newblock \bibinfo{journal}{\emph{Cancer}} \bibinfo{volume}{122}, \bibinfo{number}{13} (\bibinfo{year}{2016}), \bibinfo{pages}{1987--1995}.
\newblock


\bibitem[Khurana et~al\mbox{.}(2019)]%
        {khurana2019doctor}
\bibfield{author}{\bibinfo{person}{Sandeep Khurana}, \bibinfo{person}{Liangfei Qiu}, {and} \bibinfo{person}{Subodha Kumar}.} \bibinfo{year}{2019}\natexlab{}.
\newblock \showarticletitle{When a doctor knows, it shows: An empirical analysis of doctors’ responses in a Q\&A forum of an online healthcare portal}.
\newblock \bibinfo{journal}{\emph{Information Systems Research}} \bibinfo{volume}{30}, \bibinfo{number}{3} (\bibinfo{year}{2019}), \bibinfo{pages}{872--891}.
\newblock


\bibitem[Kim et~al\mbox{.}(2023)]%
        {kim_mindfuldiary_2023}
\bibfield{author}{\bibinfo{person}{Taewan Kim}, \bibinfo{person}{Seolyeong Bae}, \bibinfo{person}{Hyun~Ah Kim}, \bibinfo{person}{Su-woo Lee}, \bibinfo{person}{Hwajung Hong}, \bibinfo{person}{Chanmo Yang}, {and} \bibinfo{person}{Young-Ho Kim}.} \bibinfo{year}{2023}\natexlab{}.
\newblock \bibinfo{title}{{MindfulDiary}: {Harnessing} {Large} {Language} {Model} to {Support} {Psychiatric} {Patients}' {Journaling}}.
\newblock
\newblock
\urldef\tempurl%
\url{http://arxiv.org/abs/2310.05231}
\showURL{%
\tempurl}
\newblock
\shownote{arXiv:2310.05231 [cs]}.


\bibitem[Lee et~al\mbox{.}(2020)]%
        {lee2020biobert}
\bibfield{author}{\bibinfo{person}{Jinhyuk Lee}, \bibinfo{person}{Wonjin Yoon}, \bibinfo{person}{Sungdong Kim}, \bibinfo{person}{Donghyeon Kim}, \bibinfo{person}{Sunkyu Kim}, \bibinfo{person}{Chan~Ho So}, {and} \bibinfo{person}{Jaewoo Kang}.} \bibinfo{year}{2020}\natexlab{}.
\newblock \showarticletitle{BioBERT: a pre-trained biomedical language representation model for biomedical text mining}.
\newblock \bibinfo{journal}{\emph{Bioinformatics}} \bibinfo{volume}{36}, \bibinfo{number}{4} (\bibinfo{year}{2020}), \bibinfo{pages}{1234--1240}.
\newblock


\bibitem[Liu et~al\mbox{.}(2023)]%
        {liu2023utility}
\bibfield{author}{\bibinfo{person}{Jialin Liu}, \bibinfo{person}{Changyu Wang}, {and} \bibinfo{person}{Siru Liu}.} \bibinfo{year}{2023}\natexlab{}.
\newblock \showarticletitle{Utility of ChatGPT in clinical practice}.
\newblock \bibinfo{journal}{\emph{Journal of Medical Internet Research}}  \bibinfo{volume}{25} (\bibinfo{year}{2023}), \bibinfo{pages}{e48568}.
\newblock


\bibitem[Luo et~al\mbox{.}(2022)]%
        {luo2022biogpt}
\bibfield{author}{\bibinfo{person}{Renqian Luo}, \bibinfo{person}{Liai Sun}, \bibinfo{person}{Yingce Xia}, \bibinfo{person}{Tao Qin}, \bibinfo{person}{Sheng Zhang}, \bibinfo{person}{Hoifung Poon}, {and} \bibinfo{person}{Tie-Yan Liu}.} \bibinfo{year}{2022}\natexlab{}.
\newblock \showarticletitle{BioGPT: generative pre-trained transformer for biomedical text generation and mining}.
\newblock \bibinfo{journal}{\emph{Briefings in Bioinformatics}} \bibinfo{volume}{23}, \bibinfo{number}{6} (\bibinfo{year}{2022}), \bibinfo{pages}{bbac409}.
\newblock


\bibitem[Ma et~al\mbox{.}(2023)]%
        {ma2023understanding}
\bibfield{author}{\bibinfo{person}{Zilin Ma}, \bibinfo{person}{Yiyang Mei}, {and} \bibinfo{person}{Zhaoyuan Su}.} \bibinfo{year}{2023}\natexlab{}.
\newblock \showarticletitle{Understanding the benefits and challenges of using large language model-based conversational agents for mental well-being support}.
\newblock \bibinfo{journal}{\emph{arXiv preprint arXiv:2307.15810}} (\bibinfo{year}{2023}).
\newblock


\bibitem[Mahvi et~al\mbox{.}(2018)]%
        {mahvi2018local}
\bibfield{author}{\bibinfo{person}{David~A Mahvi}, \bibinfo{person}{Rong Liu}, \bibinfo{person}{Mark~W Grinstaff}, \bibinfo{person}{Yolonda~L Colson}, {and} \bibinfo{person}{Chandrajit~P Raut}.} \bibinfo{year}{2018}\natexlab{}.
\newblock \showarticletitle{Local cancer recurrence: the realities, challenges, and opportunities for new therapies}.
\newblock \bibinfo{journal}{\emph{CA: a cancer journal for clinicians}} \bibinfo{volume}{68}, \bibinfo{number}{6} (\bibinfo{year}{2018}), \bibinfo{pages}{488--505}.
\newblock


\bibitem[Masum et~al\mbox{.}(2022)]%
        {masum2022data}
\bibfield{author}{\bibinfo{person}{Shamsul Masum}, \bibinfo{person}{Adrian Hopgood}, \bibinfo{person}{Samuel Stefan}, \bibinfo{person}{Karen Flashman}, {and} \bibinfo{person}{Jim Khan}.} \bibinfo{year}{2022}\natexlab{}.
\newblock \showarticletitle{Data analytics and artificial intelligence in predicting length of stay, readmission, and mortality: a population-based study of surgical management of colorectal cancer}.
\newblock \bibinfo{journal}{\emph{Discover Oncology}} \bibinfo{volume}{13}, \bibinfo{number}{1} (\bibinfo{year}{2022}), \bibinfo{pages}{11}.
\newblock


\bibitem[Masum et~al\mbox{.}(2021)]%
        {masum2021data}
\bibfield{author}{\bibinfo{person}{Shamsul Masum}, \bibinfo{person}{Adrian Hopgood}, \bibinfo{person}{Samuel Stefan}, \bibinfo{person}{Karen Flashman}, {and} \bibinfo{person}{Jim~S Khan}.} \bibinfo{year}{2021}\natexlab{}.
\newblock \showarticletitle{Data analytics and artificial intelligence to predict length of stay, readmission and mortality after colorectal cancer surgery}.
\newblock \bibinfo{journal}{\emph{European Journal of Surgical Oncology}} \bibinfo{volume}{47}, \bibinfo{number}{2} (\bibinfo{year}{2021}), \bibinfo{pages}{e5}.
\newblock


\bibitem[Michalopoulos et~al\mbox{.}(2020)]%
        {michalopoulos_umlsbert_2020}
\bibfield{author}{\bibinfo{person}{George Michalopoulos}, \bibinfo{person}{Yuanxin Wang}, \bibinfo{person}{Hussam Kaka}, \bibinfo{person}{Helen Chen}, {and} \bibinfo{person}{Alexander Wong}.} \bibinfo{year}{2020}\natexlab{}.
\newblock \bibinfo{title}{{UmlsBERT}: {Clinical} {Domain} {Knowledge} {Augmentation} of {Contextual} {Embeddings} {Using} the {Unified} {Medical} {Language} {System} {Metathesaurus}}.
\newblock
\newblock
\urldef\tempurl%
\url{https://arxiv.org/abs/2010.10391v5}
\showURL{%
\tempurl}


\bibitem[Mohanty et~al\mbox{.}(2022)]%
        {mohanty2022machine}
\bibfield{author}{\bibinfo{person}{Somya~D Mohanty}, \bibinfo{person}{Deborah Lekan}, \bibinfo{person}{Thomas~P McCoy}, \bibinfo{person}{Marjorie Jenkins}, {and} \bibinfo{person}{Prashanti Manda}.} \bibinfo{year}{2022}\natexlab{}.
\newblock \showarticletitle{Machine learning for predicting readmission risk among the frail: Explainable AI for healthcare}.
\newblock \bibinfo{journal}{\emph{Patterns}} \bibinfo{volume}{3}, \bibinfo{number}{1} (\bibinfo{year}{2022}).
\newblock


\bibitem[Mosher et~al\mbox{.}(2016)]%
        {mosher2016mental}
\bibfield{author}{\bibinfo{person}{Catherine~E Mosher}, \bibinfo{person}{Joseph~G Winger}, \bibinfo{person}{Barbara~A Given}, \bibinfo{person}{Paul~R Helft}, {and} \bibinfo{person}{Bert~H O'Neil}.} \bibinfo{year}{2016}\natexlab{}.
\newblock \showarticletitle{Mental health outcomes during colorectal cancer survivorship: a review of the literature}.
\newblock \bibinfo{journal}{\emph{Psycho-oncology}} \bibinfo{volume}{25}, \bibinfo{number}{11} (\bibinfo{year}{2016}), \bibinfo{pages}{1261--1270}.
\newblock


\bibitem[Nahum-Shani et~al\mbox{.}(2018)]%
        {nahum2018just}
\bibfield{author}{\bibinfo{person}{Inbal Nahum-Shani}, \bibinfo{person}{Shawna~N Smith}, \bibinfo{person}{Bonnie~J Spring}, \bibinfo{person}{Linda~M Collins}, \bibinfo{person}{Katie Witkiewitz}, \bibinfo{person}{Ambuj Tewari}, {and} \bibinfo{person}{Susan~A Murphy}.} \bibinfo{year}{2018}\natexlab{}.
\newblock \showarticletitle{Just-in-time adaptive interventions (JITAIs) in mobile health: key components and design principles for ongoing health behavior support}.
\newblock \bibinfo{journal}{\emph{Annals of Behavioral Medicine}} (\bibinfo{year}{2018}), \bibinfo{pages}{1--17}.
\newblock


\bibitem[Offiah and Hall(2011)]%
        {offiah2011post}
\bibfield{author}{\bibinfo{person}{C Offiah} {and} \bibinfo{person}{E Hall}.} \bibinfo{year}{2011}\natexlab{}.
\newblock \showarticletitle{Post-treatment imaging appearances in head and neck cancer patients}.
\newblock \bibinfo{journal}{\emph{Clinical radiology}} \bibinfo{volume}{66}, \bibinfo{number}{1} (\bibinfo{year}{2011}), \bibinfo{pages}{13--24}.
\newblock


\bibitem[Parker et~al\mbox{.}(2009)]%
        {parkerBreastCancerUnique2009}
\bibfield{author}{\bibinfo{person}{Patricia~A. Parker}, \bibinfo{person}{Joann Aaron}, {and} \bibinfo{person}{Walter~F. Baile}.} \bibinfo{year}{2009}\natexlab{}.
\newblock \showarticletitle{Breast {{Cancer}}: {{Unique Communication Challenges}} and {{Strategies}} to {{Address}} Them}.
\newblock \bibinfo{journal}{\emph{The Breast Journal}} \bibinfo{volume}{15}, \bibinfo{number}{1} (\bibinfo{year}{2009}), \bibinfo{pages}{69--75}.
\newblock
\showISSN{1524-4741}
\urldef\tempurl%
\url{https://doi.org/10.1111/j.1524-4741.2008.00673.x}
\showDOI{\tempurl}


\bibitem[Penedo et~al\mbox{.}(2020)]%
        {penedo2020increasing}
\bibfield{author}{\bibinfo{person}{Frank~J Penedo}, \bibinfo{person}{Laura~B Oswald}, \bibinfo{person}{Joshua~P Kronenfeld}, \bibinfo{person}{Sofia~F Garcia}, \bibinfo{person}{David Cella}, {and} \bibinfo{person}{Betina Yanez}.} \bibinfo{year}{2020}\natexlab{}.
\newblock \showarticletitle{The increasing value of eHealth in the delivery of patient-centred cancer care}.
\newblock \bibinfo{journal}{\emph{The Lancet Oncology}} \bibinfo{volume}{21}, \bibinfo{number}{5} (\bibinfo{year}{2020}), \bibinfo{pages}{e240--e251}.
\newblock


\bibitem[Pine et~al\mbox{.}(2018)]%
        {pine_data_2018}
\bibfield{author}{\bibinfo{person}{Kathleen~H. Pine}, \bibinfo{person}{Claus Bossen}, \bibinfo{person}{Yunan Chen}, \bibinfo{person}{Gunnar Ellingsen}, \bibinfo{person}{Miria Grisot}, \bibinfo{person}{Melissa Mazmanian}, {and} \bibinfo{person}{Naja~Holten M{\o}ller}.} \bibinfo{year}{2018}\natexlab{}.
\newblock \showarticletitle{Data {Work} in {Healthcare}: {Challenges} for {Patients}, {Clinicians} and {Administrators}}. In \bibinfo{booktitle}{\emph{Companion of the 2018 {ACM} {Conference} on {Computer} {Supported} {Cooperative} {Work} and {Social} {Computing}}}. \bibinfo{publisher}{ACM}, \bibinfo{address}{Jersey City NJ USA}, \bibinfo{pages}{433--439}.
\newblock
\showISBNx{978-1-4503-6018-0}
\urldef\tempurl%
\url{https://doi.org/10.1145/3272973.3273017}
\showDOI{\tempurl}


\bibitem[Raj et~al\mbox{.}(2017)]%
        {raj_understanding_2017}
\bibfield{author}{\bibinfo{person}{Shriti Raj}, \bibinfo{person}{Mark~W. Newman}, \bibinfo{person}{Joyce~M. Lee}, {and} \bibinfo{person}{Mark~S. Ackerman}.} \bibinfo{year}{2017}\natexlab{}.
\newblock \showarticletitle{Understanding {Individual} and {Collaborative} {Problem}-{Solving} with {Patient}-{Generated} {Data}: {Challenges} and {Opportunities}}.
\newblock \bibinfo{journal}{\emph{Proceedings of the ACM on Human-Computer Interaction}} \bibinfo{volume}{1}, \bibinfo{number}{CSCW} (\bibinfo{date}{Dec.} \bibinfo{year}{2017}), \bibinfo{pages}{88:1--88:18}.
\newblock
\urldef\tempurl%
\url{https://doi.org/10.1145/3134723}
\showDOI{\tempurl}


\bibitem[Ran et~al\mbox{.}(2016)]%
        {ran_quality_2016}
\bibfield{author}{\bibinfo{person}{Lingyun Ran}, \bibinfo{person}{Xiaodong Jiang}, \bibinfo{person}{Erzhuang Qian}, \bibinfo{person}{Hongqian Kong}, \bibinfo{person}{Xiaolan Wang}, {and} \bibinfo{person}{Qin Liu}.} \bibinfo{year}{2016}\natexlab{}.
\newblock \showarticletitle{Quality of life, self-care knowledge access, and self-care needs in patients with colon stomas one month post-surgery in a {Chinese} {Tumor} {Hospital}}.
\newblock \bibinfo{journal}{\emph{International Journal of Nursing Sciences}} \bibinfo{volume}{3}, \bibinfo{number}{3} (\bibinfo{date}{Sept.} \bibinfo{year}{2016}), \bibinfo{pages}{252--258}.
\newblock
\showISSN{23520132}
\urldef\tempurl%
\url{https://doi.org/10.1016/j.ijnss.2016.07.004}
\showDOI{\tempurl}


\bibitem[Rao et~al\mbox{.}(2023)]%
        {rao_assessing_2023}
\bibfield{author}{\bibinfo{person}{Arya Rao}, \bibinfo{person}{Michael Pang}, \bibinfo{person}{John Kim}, \bibinfo{person}{Meghana Kamineni}, \bibinfo{person}{Winston Lie}, \bibinfo{person}{Anoop~K. Prasad}, \bibinfo{person}{Adam Landman}, \bibinfo{person}{Keith~J. Dreyer}, {and} \bibinfo{person}{Marc~D. Succi}.} \bibinfo{year}{2023}\natexlab{}.
\newblock \bibinfo{title}{Assessing the {Utility} of {ChatGPT} {Throughout} the {Entire} {Clinical} {Workflow}}.
\newblock
\newblock
\urldef\tempurl%
\url{https://doi.org/10.1101/2023.02.21.23285886}
\showDOI{\tempurl}
\newblock
\shownote{Pages: 2023.02.21.23285886}.


\bibitem[Rights~(OCR)(2021)]%
        {rightsocrHealthInformationPrivacy2021}
\bibfield{author}{\bibinfo{person}{Office for~Civil Rights~(OCR)}.} \bibinfo{year}{2021}\natexlab{}.
\newblock \bibinfo{title}{Health {{Information Privacy}}}.
\newblock \bibinfo{howpublished}{https://www.hhs.gov/hipaa/index.html}.
\newblock


\bibitem[Schroeder et~al\mbox{.}(2017)]%
        {schroeder_supporting_2017}
\bibfield{author}{\bibinfo{person}{Jessica Schroeder}, \bibinfo{person}{Jane Hoffswell}, \bibinfo{person}{Chia-Fang Chung}, \bibinfo{person}{James Fogarty}, \bibinfo{person}{Sean Munson}, {and} \bibinfo{person}{Jasmine Zia}.} \bibinfo{year}{2017}\natexlab{}.
\newblock \showarticletitle{Supporting {Patient}-{Provider} {Collaboration} to {Identify} {Individual} {Triggers} using {Food} and {Symptom} {Journals}}. In \bibinfo{booktitle}{\emph{Proceedings of the 2017 {ACM} {Conference} on {Computer} {Supported} {Cooperative} {Work} and {Social} {Computing}}} \emph{(\bibinfo{series}{{CSCW} '17})}. \bibinfo{publisher}{Association for Computing Machinery}, \bibinfo{address}{New York, NY, USA}, \bibinfo{pages}{1726--1739}.
\newblock
\showISBNx{978-1-4503-4335-0}
\urldef\tempurl%
\url{https://doi.org/10.1145/2998181.2998276}
\showDOI{\tempurl}


\bibitem[Seo et~al\mbox{.}(2021)]%
        {seo2021challenges}
\bibfield{author}{\bibinfo{person}{Woosuk Seo}, \bibinfo{person}{Ayse~G Buyuktur}, \bibinfo{person}{Sung~Won Choi}, \bibinfo{person}{Laura Sedig}, {and} \bibinfo{person}{Sun~Young Park}.} \bibinfo{year}{2021}\natexlab{}.
\newblock \showarticletitle{Challenges in the parent-child communication of health-related information in pediatric cancer care}.
\newblock \bibinfo{journal}{\emph{Proceedings of the ACM on Human-Computer Interaction}} \bibinfo{volume}{5}, \bibinfo{number}{CSCW1} (\bibinfo{year}{2021}), \bibinfo{pages}{1--24}.
\newblock


\bibitem[Seo et~al\mbox{.}(2023)]%
        {seo_chacha_2023}
\bibfield{author}{\bibinfo{person}{Woosuk Seo}, \bibinfo{person}{Chanmo Yang}, {and} \bibinfo{person}{Young-Ho Kim}.} \bibinfo{year}{2023}\natexlab{}.
\newblock \bibinfo{title}{{ChaCha}: {Leveraging} {Large} {Language} {Models} to {Prompt} {Children} to {Share} {Their} {Emotions} about {Personal} {Events}}.
\newblock
\newblock
\urldef\tempurl%
\url{http://arxiv.org/abs/2309.12244}
\showURL{%
\tempurl}
\newblock
\shownote{arXiv:2309.12244 [cs]}.


\bibitem[Serban et~al\mbox{.}(2023)]%
        {serban_i_2023}
\bibfield{author}{\bibinfo{person}{Irina~Bianca Serban}, \bibinfo{person}{Dimitra Dritsa}, \bibinfo{person}{Israel Campero~Jurado}, \bibinfo{person}{Steven Houben}, \bibinfo{person}{Aarnout Brombacher}, \bibinfo{person}{David Ten~Cate}, \bibinfo{person}{Loes Janssen}, {and} \bibinfo{person}{Margot Heijmans}.} \bibinfo{year}{2023}\natexlab{}.
\newblock \showarticletitle{``{I} just see numbers, but how do you feel about your training?'': {Clinicians}' {Data} {Needs} in {Telemonitoring} for {Colorectal} {Cancer} {Surgery} {Prehabilitation}}. In \bibinfo{booktitle}{\emph{Computer {Supported} {Cooperative} {Work} and {Social} {Computing}}}. \bibinfo{publisher}{ACM}, \bibinfo{address}{Minneapolis MN USA}, \bibinfo{pages}{267--272}.
\newblock
\showISBNx{9798400701290}
\urldef\tempurl%
\url{https://doi.org/10.1145/3584931.3607006}
\showDOI{\tempurl}


\bibitem[Siegel et~al\mbox{.}(2021)]%
        {siegel2021cancer}
\bibfield{author}{\bibinfo{person}{Rebecca~L Siegel}, \bibinfo{person}{Kimberly~D Miller}, \bibinfo{person}{Hannah~E Fuchs}, \bibinfo{person}{Ahmedin Jemal}, {et~al\mbox{.}}} \bibinfo{year}{2021}\natexlab{}.
\newblock \showarticletitle{Cancer statistics, 2021}.
\newblock \bibinfo{journal}{\emph{Ca Cancer J Clin}} \bibinfo{volume}{71}, \bibinfo{number}{1} (\bibinfo{year}{2021}), \bibinfo{pages}{7--33}.
\newblock


\bibitem[Silva et~al\mbox{.}(2020)]%
        {silva_ostomy_2020}
\bibfield{author}{\bibinfo{person}{Karine De~Almeida Silva}, \bibinfo{person}{Arenamoline~Xavier Duarte}, \bibinfo{person}{Amanda~Rodrigues Cruz}, \bibinfo{person}{Letícia~Oliveira Cardoso}, \bibinfo{person}{Thatty Christina~Morais Santos}, {and} \bibinfo{person}{Geórgia Das~Graças Pena}.} \bibinfo{year}{2020}\natexlab{}.
\newblock \showarticletitle{Ostomy time and nutrition status were associated on quality of life in patients with colorectal cancer}.
\newblock \bibinfo{journal}{\emph{Journal of Coloproctology}} \bibinfo{volume}{40}, \bibinfo{number}{04} (\bibinfo{date}{Dec.} \bibinfo{year}{2020}), \bibinfo{pages}{352--361}.
\newblock
\showISSN{2237-9363, 2317-6423}
\urldef\tempurl%
\url{https://doi.org/10.1016/j.jcol.2020.07.003}
\showDOI{\tempurl}


\bibitem[Subramani et~al\mbox{.}(2023)]%
        {subramani2023detecting}
\bibfield{author}{\bibinfo{person}{Nishant Subramani}, \bibinfo{person}{Sasha Luccioni}, \bibinfo{person}{Jesse Dodge}, {and} \bibinfo{person}{Margaret Mitchell}.} \bibinfo{year}{2023}\natexlab{}.
\newblock \showarticletitle{Detecting personal information in training corpora: an analysis}. In \bibinfo{booktitle}{\emph{Proceedings of the 3rd Workshop on Trustworthy Natural Language Processing (TrustNLP 2023)}}. \bibinfo{pages}{208--220}.
\newblock


\bibitem[Suh et~al\mbox{.}(2020)]%
        {suh_parallel_2020}
\bibfield{author}{\bibinfo{person}{Jina Suh}, \bibinfo{person}{Spencer Williams}, \bibinfo{person}{Jesse~R. Fann}, \bibinfo{person}{James Fogarty}, \bibinfo{person}{Amy~M. Bauer}, {and} \bibinfo{person}{Gary Hsieh}.} \bibinfo{year}{2020}\natexlab{}.
\newblock \showarticletitle{Parallel {Journeys} of {Patients} with {Cancer} and {Depression}: {Challenges} and {Opportunities} for {Technology}-{Enabled} {Collaborative} {Care}}.
\newblock \bibinfo{journal}{\emph{Proceedings of the ACM on Human-Computer Interaction}} \bibinfo{volume}{4}, \bibinfo{number}{CSCW1} (\bibinfo{date}{May} \bibinfo{year}{2020}), \bibinfo{pages}{1--36}.
\newblock
\showISSN{2573-0142}
\urldef\tempurl%
\url{https://doi.org/10.1145/3392843}
\showDOI{\tempurl}


\bibitem[Surbone et~al\mbox{.}(2012)]%
        {surboneNewChallengesCommunication2012}
\bibfield{author}{\bibinfo{person}{Antonella Surbone}, \bibinfo{person}{Matja{\v z} Zwitter}, \bibinfo{person}{Mirjana Rajer}, {and} \bibinfo{person}{Richard Stiefel}.} \bibinfo{year}{2012}\natexlab{}.
\newblock \bibinfo{booktitle}{\emph{New {{Challenges}} in {{Communication}} with {{Cancer Patients}}}}.
\newblock \bibinfo{publisher}{{Springer Science \& Business Media}}.
\newblock
\showISBNx{978-1-4614-3369-9}


\bibitem[Tang et~al\mbox{.}(2023)]%
        {tang_does_2023}
\bibfield{author}{\bibinfo{person}{Ruixiang Tang}, \bibinfo{person}{Xiaotian Han}, \bibinfo{person}{Xiaoqian Jiang}, {and} \bibinfo{person}{Xia Hu}.} \bibinfo{year}{2023}\natexlab{}.
\newblock \bibinfo{title}{Does {Synthetic} {Data} {Generation} of {LLMs} {Help} {Clinical} {Text} {Mining}?}
\newblock
\newblock
\urldef\tempurl%
\url{http://arxiv.org/abs/2303.04360}
\showURL{%
\tempurl}
\newblock
\shownote{arXiv:2303.04360 [cs]}.


\bibitem[Thorne et~al\mbox{.}(2013)]%
        {thorne2013communication}
\bibfield{author}{\bibinfo{person}{Sally~E Thorne}, \bibinfo{person}{John~L Oliffe}, \bibinfo{person}{Valerie Oglov}, {and} \bibinfo{person}{Karen Gelmon}.} \bibinfo{year}{2013}\natexlab{}.
\newblock \showarticletitle{Communication challenges for chronic metastatic cancer in an era of novel therapeutics}.
\newblock \bibinfo{journal}{\emph{Qualitative health research}} \bibinfo{volume}{23}, \bibinfo{number}{7} (\bibinfo{year}{2013}), \bibinfo{pages}{863--875}.
\newblock


\bibitem[Vachon(1995)]%
        {vachon1995staff}
\bibfield{author}{\bibinfo{person}{Mary~LS Vachon}.} \bibinfo{year}{1995}\natexlab{}.
\newblock \showarticletitle{Staff stress in hospice/palliative care: a review}.
\newblock \bibinfo{journal}{\emph{Palliative medicine}} \bibinfo{volume}{9}, \bibinfo{number}{2} (\bibinfo{year}{1995}), \bibinfo{pages}{91--122}.
\newblock


\bibitem[Wang et~al\mbox{.}(2020)]%
        {wang_please_2020}
\bibfield{author}{\bibinfo{person}{Ding Wang}, \bibinfo{person}{Santosh~D. Kale}, {and} \bibinfo{person}{Jacki O'Neill}.} \bibinfo{year}{2020}\natexlab{}.
\newblock \showarticletitle{Please {Call} the {Specialism}: {Using} {WeChat} to {Support} {Patient} {Care} in {China}}. In \bibinfo{booktitle}{\emph{Proceedings of the 2020 {CHI} {Conference} on {Human} {Factors} in {Computing} {Systems}}}. \bibinfo{publisher}{ACM}, \bibinfo{address}{Honolulu HI USA}, \bibinfo{pages}{1--13}.
\newblock
\showISBNx{978-1-4503-6708-0}
\urldef\tempurl%
\url{https://doi.org/10.1145/3313831.3376274}
\showDOI{\tempurl}


\bibitem[Wang et~al\mbox{.}(2018)]%
        {wang2018prevalence}
\bibfield{author}{\bibinfo{person}{Katie Wang}, \bibinfo{person}{Caitlin Yee}, \bibinfo{person}{Samantha Tam}, \bibinfo{person}{Leah Drost}, \bibinfo{person}{Stephanie Chan}, \bibinfo{person}{Pearl Zaki}, \bibinfo{person}{Victoria Rico}, \bibinfo{person}{Krista Ariello}, \bibinfo{person}{Mark Dasios}, \bibinfo{person}{Henry Lam}, {et~al\mbox{.}}} \bibinfo{year}{2018}\natexlab{}.
\newblock \showarticletitle{Prevalence of pain in patients with breast cancer post-treatment: A systematic review}.
\newblock \bibinfo{journal}{\emph{The Breast}}  \bibinfo{volume}{42} (\bibinfo{year}{2018}), \bibinfo{pages}{113--127}.
\newblock


\bibitem[Wang and Miller(2020)]%
        {wang2020just}
\bibfield{author}{\bibinfo{person}{Liyuan Wang} {and} \bibinfo{person}{Lynn~Carol Miller}.} \bibinfo{year}{2020}\natexlab{}.
\newblock \showarticletitle{Just-in-the-moment adaptive interventions (JITAI): A meta-analytical review}.
\newblock \bibinfo{journal}{\emph{Health Communication}} \bibinfo{volume}{35}, \bibinfo{number}{12} (\bibinfo{year}{2020}), \bibinfo{pages}{1531--1544}.
\newblock


\bibitem[Wu et~al\mbox{.}(2023)]%
        {wu_mindshift_2023}
\bibfield{author}{\bibinfo{person}{Ruolan Wu}, \bibinfo{person}{Chun Yu}, \bibinfo{person}{Xiaole Pan}, \bibinfo{person}{Yujia Liu}, \bibinfo{person}{Ningning Zhang}, \bibinfo{person}{Yue Fu}, \bibinfo{person}{Yuhan Wang}, \bibinfo{person}{Zhi Zheng}, \bibinfo{person}{Li Chen}, \bibinfo{person}{Qiaolei Jiang}, \bibinfo{person}{Xuhai Xu}, {and} \bibinfo{person}{Yuanchun Shi}.} \bibinfo{year}{2023}\natexlab{}.
\newblock \bibinfo{title}{{MindShift}: {Leveraging} {Large} {Language} {Models} for {Mental}-{States}-{Based} {Problematic} {Smartphone} {Use} {Intervention}}.
\newblock
\newblock
\urldef\tempurl%
\url{http://arxiv.org/abs/2309.16639}
\showURL{%
\tempurl}
\newblock
\shownote{arXiv:2309.16639 [cs]}.


\bibitem[Xu et~al\mbox{.}(2023)]%
        {xu2023leveraging}
\bibfield{author}{\bibinfo{person}{Xuhai Xu}, \bibinfo{person}{Bingshen Yao}, \bibinfo{person}{Yuanzhe Dong}, \bibinfo{person}{Hong Yu}, \bibinfo{person}{James Hendler}, \bibinfo{person}{Anind~K Dey}, {and} \bibinfo{person}{Dakuo Wang}.} \bibinfo{year}{2023}\natexlab{}.
\newblock \showarticletitle{Leveraging large language models for mental health prediction via online text data}.
\newblock \bibinfo{journal}{\emph{arXiv preprint arXiv:2307.14385}} (\bibinfo{year}{2023}).
\newblock


\bibitem[Yang et~al\mbox{.}(2023a)]%
        {yang_integrating_2023}
\bibfield{author}{\bibinfo{person}{Rui Yang}, \bibinfo{person}{Edison Marrese-Taylor}, \bibinfo{person}{Yuhe Ke}, \bibinfo{person}{Lechao Cheng}, \bibinfo{person}{Qingyu Chen}, {and} \bibinfo{person}{Irene Li}.} \bibinfo{year}{2023}\natexlab{a}.
\newblock \bibinfo{title}{Integrating {UMLS} {Knowledge} into {Large} {Language} {Models} for {Medical} {Question} {Answering}}.
\newblock
\newblock
\urldef\tempurl%
\url{http://arxiv.org/abs/2310.02778}
\showURL{%
\tempurl}
\newblock
\shownote{arXiv:2310.02778 [cs]}.


\bibitem[Yang et~al\mbox{.}(2023b)]%
        {yang2023talk2care}
\bibfield{author}{\bibinfo{person}{Ziqi Yang}, \bibinfo{person}{Xuhai Xu}, \bibinfo{person}{Bingsheng Yao}, \bibinfo{person}{Shao Zhang}, \bibinfo{person}{Ethan Rogers}, \bibinfo{person}{Stephen Intille}, \bibinfo{person}{Nawar Shara}, \bibinfo{person}{Dakuo Wang}, {et~al\mbox{.}}} \bibinfo{year}{2023}\natexlab{b}.
\newblock \showarticletitle{Talk2Care: Facilitating Asynchronous Patient-Provider Communication with Large-Language-Model}.
\newblock \bibinfo{journal}{\emph{arXiv preprint arXiv:2309.09357}} (\bibinfo{year}{2023}).
\newblock


\bibitem[Ye et~al\mbox{.}(2010)]%
        {ye_e-mail_2010}
\bibfield{author}{\bibinfo{person}{Jiali Ye}, \bibinfo{person}{George Rust}, \bibinfo{person}{Yvonne Fry-Johnson}, {and} \bibinfo{person}{Harry Strothers}.} \bibinfo{year}{2010}\natexlab{}.
\newblock \showarticletitle{E-mail in patient--provider communication: {A} systematic review}.
\newblock \bibinfo{journal}{\emph{Patient Education and Counseling}} \bibinfo{volume}{80}, \bibinfo{number}{2} (\bibinfo{date}{Aug.} \bibinfo{year}{2010}), \bibinfo{pages}{266--273}.
\newblock
\showISSN{07383991}
\urldef\tempurl%
\url{https://doi.org/10.1016/j.pec.2009.09.038}
\showDOI{\tempurl}


\bibitem[Yu et~al\mbox{.}(2021)]%
        {yu2021deep}
\bibfield{author}{\bibinfo{person}{Keping Yu}, \bibinfo{person}{Liang Tan}, \bibinfo{person}{Long Lin}, \bibinfo{person}{Xiaofan Cheng}, \bibinfo{person}{Zhang Yi}, {and} \bibinfo{person}{Takuro Sato}.} \bibinfo{year}{2021}\natexlab{}.
\newblock \showarticletitle{Deep-learning-empowered breast cancer auxiliary diagnosis for 5GB remote E-health}.
\newblock \bibinfo{journal}{\emph{IEEE Wireless Communications}} \bibinfo{volume}{28}, \bibinfo{number}{3} (\bibinfo{year}{2021}), \bibinfo{pages}{54--61}.
\newblock


\end{thebibliography}

%%
%% If your work has an appendix, this is the place to put it.
% \appendix
% \input{7-appendix}

\end{document}